\documentclass[iop,apj]{emulateapj}
 
\usepackage{epstopdf}
\usepackage{epsfig}

\def\deg      {{\ifmmode^\circ\else$^\circ$\fi}} 

\defcitealias{Wyder2007}{Paper I} 
\defcitealias{Martin2007}{Paper III} 
\defcitealias{Salim2007}{S07}  
 
\slugcomment{ApJ, in press}
 
\begin{document}

\bibliographystyle{apj}

\defcitealias{Thilker07}{T07}


\title{The Space Density of Extended Ultraviolet (XUV) disks in the Local Universe and Implications for Gas Accretion onto Galaxies}

 

%
%
%
\author{
Jenna J. Lemonias\altaffilmark{1},
David Schiminovich\altaffilmark{1},  
David Thilker\altaffilmark{2},
Ted K. Wyder\altaffilmark{3}, 
D. Christopher Martin\altaffilmark{3}, 
Mark Seibert\altaffilmark{4}, 
Marie A. Treyer\altaffilmark{5}, 
Luciana Bianchi\altaffilmark{2}, 
Timothy M. Heckman\altaffilmark{6},
Barry F. Madore\altaffilmark{4}, 
R. Michael Rich\altaffilmark{7}
}

\altaffiltext{1}{Department of Astronomy, Columbia University, 550 West 120th Street, New York, NY 10027; jenna@astro.columbia.edu}
\altaffiltext{2}{Center for Astrophysical Sciences, The Johns Hopkins` University, 3400 N. Charles St., Baltimore, MD 21218}
\altaffiltext{3}{California Institute of Technology, MC 278-17, 1200 East California Boulevard, Pasadena, CA 91125}
\altaffiltext{4}{Observatories of the Carnegie Institution of Washington, 813 Santa Barbara St., Pasadena, CA 91101}
\altaffiltext{5}{Laboratoire d'Astrophysique de Marseille, BP8, Traverse du Siphon, F-13376 Marseille, France}
\altaffiltext{6}{Department of Physics and Astronomy, The Johns Hopkins University, Homewood Campus, Baltimore, MD 21218}
\altaffiltext{7}{Department of Physics and Astronomy, University of California, Los Angeles, CA 90095}

 
%

%
 
  
\begin{abstract}

We present results of the first unbiased search for extended UV (XUV)-disk galaxies undertaken to determine the space density of such galaxies.  Our sample contains 561 local (0.001 $<$ z $<$ 0.05) galaxies that lie in the intersection of available \emph{GALEX} deep imaging (exposure time $>$ 1.5 $\times$ 10$^{4}$ s) and SDSS DR7 footprints.   We explore modifications to the standard classification scheme for our sample that includes both disk- and bulge-dominated galaxies.  Visual classification of each galaxy in the sample reveals an XUV-disk frequency of up to 20\% for the most nearby portion of our sample.  On average over the entire sample (out to z=0.05) the frequency ranges from a hard limit of 4\% to 14\%.   The \emph{GALEX} imaging allows us to detect XUV-disks beyond 100 Mpc.  The XUV regions around XUV-disk galaxies are consistently bluer than the main bodies.  We find a surprisingly high frequency of XUV emission around luminous red (NUV-\emph{r} $>$ 5) and green valley (3 $<$ NUV-\emph{r} $<$ 5) galaxies.  The XUV-disk space density in the local universe is $>$ 1.5-4.2 $\times$ 10$^{-3}$ Mpc$^{-3}$.  Using the XUV emission as an indicator of recent gas accretion, we estimate that the cold gas accretion rate onto these galaxies is $>$ 1.7-4.6 $\times$ 10$^{-3}$ M$_{\odot}$ Mpc$^{-3}$ yr$^{-1}$.  The number of XUV-disks in the green valley and the estimated accretion rate onto such galaxies points to the intriguing possibility that 7-18\% of galaxies in this population are transitioning away from the red sequence.

\end{abstract}

 
\keywords{galaxies: evolution --- galaxies: formation --- galaxies: structure --- surveys --- ultraviolet: galaxies }
 


\section{INTRODUCTION}  
\label{sec:intro}

Ultraviolet observations of galaxies acquired by \emph{GALEX} have recently challenged the notion of a well-defined star formation threshold by showing that outer star formation can exist at radii that are four times beyond the optical extent of the galaxy \citep{GilDePaz05, Thilker05, Thilker2010}.  Following the discovery of extended ultraviolet (XUV) emission in the outer disks of NGC 4625 and M 83 \citep{GilDePaz05, Thilker05}, Thilker et al. (2007; hereafter T07) conducted the first survey of extended UV emission in a sample of 189 nearby (d $<$ 40 Mpc) disk galaxies and concluded that 29\% of disk galaxies exhibit XUV emission.  They developed a classification system in which Type 1 XUV-disks show a number of UV-bright, optically-faint, structured complexes beyond the star formation threshold and Type 2 XUV-disks have a large, blue, low surface brightness (LSB) zone beyond the limit within which optical light dominates.

Perhaps more surprising than the existence of star formation in the outer regions of disks is the existence of XUV-disks around early-type galaxies.  In a survey of 31 E/S0 galaxies, \citet{Moffett2010b} identified 13 as XUV-disks.  NGC 404 \citep{Thilker2010} and ESO 381-47 \citep{Donovan09} are both early-type galaxies considered Type 1 XUV-disks because of a ring of star formation discernible in the UV.  \citet{Salim2010} revealed a population of massive early-type galaxies with UV excess that contain extended UV structures.  XUV-disks around early-type galaxies are especially puzzling because they suggest that extended star formation does not require the presence of an inner star-forming disk.

A distinct but related phenomenon presented in \citet{Werk2010} is the presence of outlying HII regions around gas-rich galaxies, some of which are associated with Type 1 XUV-disks.  Out of the seven galaxies they found to be supporting outer HII regions, six of them are undergoing interactions or have nearby companions, providing evidence for interactions as a cause of extended star formation.  They estimate that 6-10\% of gas-rich galaxies contain recent star formation in their outer disks.

The existence and frequency of galaxies with XUV emission has implications regarding not just the expected radial extent of star formation but also the viability of star formation in regions of low gas density, the process of disk-building, and the causes of star formation in the outer regions of galaxies of all types.  Interactions, perturbations, gas accretion, and the outward propagation of spiral density waves have all been proposed as triggers of extended star formation \citep[T07;][]{Bush08,Bush2010}.   Recent cosmological simulations done by \citet{Roskar2010} established a possible link between XUV-disks and cold gas accretion.  A systematic analysis of the types of galaxies in which XUV-disks predominate could provide crucial information about the origin of XUV-disks and their context in the overall framework of galaxy evolution.

In this paper we extend the sample of XUV-disks to higher redshifts by conducting an unbiased survey of over 500 local galaxies with redshifts up to z=0.05.  Such a large and volume-limited sample allows us to determine the space density of XUV-disk galaxies.  We analyze the properties of the XUV-disk galaxies in relation to the rest of the sample in order to determine the types of galaxies that tend to exhibit XUV emission, the nature of the XUV emission itself, and the causes of XUV emission. We pursue the connection between XUV-disks and gas accretion posited by \citet{Roskar2010} by assuming that the XUV emission in our sample of galaxies is indicative of recent gas accretion and then calculating the expected gas accretion rate onto XUV-disks.  As part of this analysis, we build on a large, homogeneous dataset of galaxies observed by SDSS and \emph{GALEX} that includes measurements of the bivariate luminosity distribution \citep{Wyder07}.

In Section \ref{sec:sample} we describe the sample and the observations used for the survey.  Section \ref{sec:XUV-disk identification} describes the method we employed to identify XUV-disk galaxies and includes several examples of the application of our method.  In Section \ref{sec:results} we analyze the properties of the XUV-disk galaxies in relation to the rest of the sample.  Section \ref{sec:discussion} presents a discussion of our findings, including our estimate of the XUV-disk space density and a calculation of the inferred gas accretion rates onto XUV-disk galaxies.  Throughout the paper, we use the standard cosmological parameters with \emph{H$_o$} = 70 km s$^{-1}$ Mpc$^{-1}$.

\section{SAMPLE}
\label{sec:sample}

We created our sample by compiling all the \emph{GALEX} images of known galaxies in the redshift range 0.001 $< z <$ 0.05 that lie within the intersection of available \emph{GALEX} deep imaging and the SDSS DR7 footprint.  The redshift range was chosen to overlap with the redshift range of the galaxies in the low-redshift portion of the NYU Value-Added Galaxy Catalog \citep{Blanton05b, Blanton05a} containing galaxies in the range 14.3 $<$ d $<$ 214.3 Mpc.  Each galaxy was matched to the nearest primary object in the SDSS DR7 photometric sample using a 5$\arcsec$ search radius.  We began with 907 GALEX images.  Only 771 of these images were centered on unique galaxies; some galaxies had been imaged by \emph{GALEX} more than once or as part of different surveys.  

We eliminated all images from the sample in which the target galaxy was within 2$\arcmin$ of the edge or off the edge of the \emph{GALEX} detector (N=246) and all images with a bright star or \emph{GALEX} artifact close enough to the galaxy (within 30$\arcsec$) such that UV photometry would be compromised (N=6).  We also removed from the sample one image in which the galaxy was off-center in the image (N=1).  We removed galaxies with no matches to the MPA-JHU SDSS catalogs (N=8), from which we obtained optical photometry.  We restricted our sample to galaxies with 0 $<$ NUV-\emph{r} $<$ 7 and 8 $<$ log(M$_*$/M$_{\odot}$) $<$ 12 to match the sample of galaxies in Wyder et al. (2007); galaxies with values beyond these limits were removed from the sample (N=30).  We note that such extreme colors are probably not real.  After removing galaxies from the sample for the reasons mentioned above, there were still some cases in which there was more than one \emph{GALEX} image of the same galaxy (N=55); we kept only the image with the longest \emph{GALEX} exposure time.  Our final sample consists of 561 galaxies.  We present the sample in Table 1.  

\begin{deluxetable*}{lrrrcccccccccc}  
\tabletypesize{\scriptsize}
\tablecolumns{11}
\tablecaption{Properties of Sample Galaxies}
\tablewidth{0pt}
\tablehead{
\colhead{SDSS ID} & \colhead{RA} & \colhead{DEC} & \colhead{z} & 
\colhead{t$_{\mathrm{exp}}$(NUV)\tablenotemark{a}} & \colhead{t$_{\mathrm{exp}}$(FUV)\tablenotemark{b}} & \colhead{NUV-\emph{r}} & \colhead{FUV-\emph{r}} & \colhead{\emph{r}} & \colhead{log(M$_{*}$/M$_{\odot}$)} & 
\colhead{R$_{90}$/R$_{50}$} &  \\
\colhead{(1)} & \colhead{(2)} & \colhead{(3)} & \colhead{(4)} & \colhead{(5)} &
\colhead{(6)} & \colhead{(7)} & \colhead{(8)} & \colhead{(9)} & \colhead{(10)} &
\colhead{(11)} 
}
\startdata

SDSS J022731.64+005238.6	&36.8818	&0.8774	&0.041	&  30.0	&  28.2	&2.34	&2.68	&16.42	&9.67	&2.42\\
SDSS J120121.90-032201.7	&180.3413	&-3.3672	&0.019	&  20.2	&  16.7	&3.35	&5.42	&15.36	&9.95	&2.30\\
SDSS J120252.21-032603.9	&180.7176	&-3.4344	&0.047	&  20.2	&  16.7	&4.49	&5.14	&16.52	&10.51	&2.39\\
SDSS J120058.38-034116.3	&180.2433	&-3.6879	&0.026	&  20.2	&  16.7	&2.23	&2.47	&16.49	&9.05	&2.23\\
SDSS J120123.26-034100.2	&180.3469	&-3.6834	&0.026	&  20.2	&  16.7	&2.59	&3.08	&17.50	&9.10	&2.59\\
SDSS J120131.20-033702.0	&180.3800	&-3.6172	&0.027	&  20.2	&  16.7	&2.23	&2.62	&15.89	&9.48	&2.27\\
SDSS J120317.19-033949.7	&180.8217	&-3.6638	&0.042	&  20.2	&  16.7	&4.81	&5.18	&16.97	&10.26	&2.74\\
SDSS J120351.08-034234.5	&180.9628	&-3.7096	&0.013	&  20.2	&  16.7	&2.20	&2.42	&15.64	&8.67	&2.73\\
SDSS J171336.65+591839.2	&258.4027	&59.3109	&0.029	&  39.8	&  29.6	&2.43	&2.65	&17.17	&8.92	&2.14\\
SDSS J171409.03+584906.4	&258.5376	&58.8185	&0.030	&  39.8	&  29.6	&3.24	&3.69	&14.25	&10.61	&1.92\\

\enddata
\label{tbl-1}
\tablecomments{This table is available in its entirety in a machine-readable form in the online journal.  A portion is 
shown here for guidance regarding its form and content.}
\tablenotetext{a}{NUV exposure time in kiloseconds.}
\tablenotetext{b}{FUV exposure time in kiloseconds.}
\end{deluxetable*}

\subsection{\emph{GALEX} Imaging and Data}
\label{sec:GALEX}

Observations of each object were obtained from \emph{GALEX} deep imaging collected as part of primary mission surveys.  All objects were imaged by \emph{GALEX} in the near-UV (NUV; 1750-2750 \AA) and far-UV (FUV; 1350-1750 \AA) bands for a minimum of 1.5 $\times$ 10$^4$ s.  The median exposure time for galaxies in the sample is 3.7 $\times$ 10$^4$ s.  The \emph{GALEX} full-width at half-maximum (FWHM) of the point-spread function (PSF) is 5.3$\arcsec$ in the NUV and 4.2$\arcsec$ in the FUV \citep{Morrissey07}.

NUV and FUV Petrosian magnitudes \citep[as described in][]{Blanton01,Yasuda01} were computed using a custom code in which nearby objects are masked.  We used the NUV image to determine the Petrosian radius, which we then used to define the photometric aperture for the NUV and FUV images.  A standard circular aperture 15$\arcsec$ in diameter was used for small LSB galaxies.  All magnitudes cited in this paper are corrected for Galactic extinction due to dust according to \citet{Wyder07}.  

Colors of the extended UV emission within the XUV region were computed by obtaining photometry outside of a surface brightness threshold defined by  $\mu_{FUV}$=27.25 ABmag arcsec$^{-2}$ and inside the contour at which $\mu_{FUV}$=29.0 ABmag arcsec$^{-2}$.  See section \ref{sec:XUVregion} for an explanation of the XUV region.

\subsection{SDSS Imaging and Data}
\label{SDSS}

Cutouts of the SDSS \emph{r}-band FITS images and SDSS \emph{gri} color composite images of each \emph{GALEX}-SDSS matched object were generated using \emph{smosaic}, a tool developed for use with the NYU Value-Added Galaxy Catalog (Blanton et al. 2005a).  We used the \emph{r}-band FITS images to measure the optical magnitude of the extended UV emission, the limits of which are given above and explained in section \ref{sec:XUVregion}.  Optical measurements (Petrosian magnitudes, R$_{50}$, R$_{90}$, and Galactic reddening) as well as stellar mass for each object were obtained from the MPA-JHU SDSS catalogs\footnote{http://www.mpa-garching.mpg.de/SDSS/DR7/}.  Isophotal measurements that quantify the ellipticity of the galaxy (major axis, minor axis, and position angle) were obtained for purposes of photometry from the SDSS SkyServer\footnote{http://cas.sdss.org/dr7/en/tools/crossid/crossid.asp}. 

\section{XUV-DISK IDENTIFICATION}
\label{sec:XUV-disk identification}

\subsection{Methodology}
\label{sec:methodology}

\subsubsection{Previous Classification Schemes Used}
\label{sec:previous schemes}

To date, there has been only one other systematic search for XUV-disk galaxies, done by \citetalias{Thilker07}.  They identified two distinct populations of XUV-disk galaxies in a nearby (d $<$ 40 Mpc) sample of late S0s through Sm galaxies with inclination $\le$ 80$^{\circ}$.  In their classification system, Type 1 XUV-disks contain structured UV-bright emission complexes beyond the expected extent of star formation.  The UV image of each galaxy was checked against an \emph{r}-band image to verify that the UV structures are not obvious at optical wavelengths.  Type 2 XUV-disks were quantitatively identified based on the color and size of a LSB region within the star formation threshold but beyond a contour enclosing 80\% of the K-band luminosity of the galaxy.  Thus, the identification of Type 2 XUV-disks was based not on an assumed star formation threshold but on a significant level of recent widespread star formation compared to the underlying disk.  Galaxies which satisfy the requirements of both XUV-disk definitions are called ``mixed-type" XUV-disks.

In their study of XUV-disks around 30 E/S0s, \citet{Moffett09} apply the classification scheme of \citetalias{Thilker07} to a fraction of their sample but find that the locations of the two contours used to classify Type 2 XUV-disks were often interchanged, making it impossible to adhere to the classification scheme as it was designed.  We address this point in section \ref{sec:color cut}.

\subsubsection{Our Classification Scheme}
\label{sec:our scheme}

We adopted the criterion most central to \citetalias{Thilker07}'s Type 1 classification scheme, namely that there be UV emission beyond a putative star formation threshold given by $\Sigma_{SFR}$ = 3 $\times$ 10$^{-4}$ M$_\odot$ yr$^{-1}$ kpc$^{-2}$, which corresponds to $\mu_{FUV}$=27.25 ABmag arcsec$^{-2}$ using the star formation rate (SFR) calibration of \citet{Kennicutt98} and correcting for Galactic extinction.   Although we use UV flux beyond this isophote as our main initial criterion, we emphasize that we do not make any attempt to classify the galaxies in our sample as Type 1 or Type 2 XUV-disks and that the galaxies we do identify as XUV-disks do not necessarily belong in either of T07's categories.  

The theoretical and observational motivation behind the chosen star formation threshold is described in detail in T07.  We note that the primary reason for the choice of the specific UV threshold given above stems from \citet{Boissier07} which showed that the threshold given above coincides with the H$\alpha$ break found in \citet{Martin01} galaxies.  More recent work has provided further evidence of the existence of a star formation threshold by comparing the star formation rate surface density to the gas surface density and demonstrating that there is a downturn in star formation at low gas surface densities \citep{Wyder09,  Bigiel2010}.

We proceeded by overlaying the contour representing the FUV surface brightness limit, given above, on a 4$\arcmin$ $\times$ 4$\arcmin$ \emph{GALEX} FUV and NUV composite image of each galaxy.  We visually inspected each image to identify UV emission beyond that contour.  The exposure times in our survey allow us to detect LSB features down to a limiting magnitude of 29.25 ABmag arcsec$^{-2}$ with a signal to noise ratio equal to 5 on small scales comparable to the \emph{GALEX} PSF.  

Each galaxy with detected UV flux - structured or diffuse, clearly connected to the target galaxy or not - beyond the threshold was identified as an XUV-disk galaxy.  We also identified a number of galaxies as XUV-ambiguous.  (All XUV-disk and XUV-ambiguous galaxy identifications were agreed upon by J.J.L. and D.S.)  The XUV-ambiguous galaxies have UV emission just beyond the expected star formation contour that may be residual flux associated with the main star-forming disk as opposed to real extended star formation independent of the central disk.  In some cases, the UV emission that qualified a galaxy for XUV-ambiguous status may actually be a background source or part of a broad wing of the \emph{GALEX} PSF.  Nevertheless, some of the XUV-ambiguous galaxies could very well turn out to be XUV-disks when deeper or more highly resolved images are obtained.  In the following analysis, we include both the definite XUV-disks and the XUV-ambiguous galaxies to provide some measure of the uncertainty in our classification procedure though we treat them separately at times to point out the differences between the two populations.

Here, we do not impose any restrictions on the optical brightness of the XUV-disk or its component structures.  In section \ref{sec:color cut} and the Appendix we examine how such a restriction may change our sample and conclusions.

\subsubsection{Limits of our Classification Scheme}
\label{sec:limits}

We recognize that our classification scheme is less quantitative and therefore more ambiguous than previous XUV-disk classification schemes.  The ambiguity inherent in our visual classification scheme may cause us to identify XUV-disks where others may not have seen XUV-disks, and it is quite likely that the limits imposed on our study by the resolution of \emph{GALEX} causes us to underestimate the frequency of XUV-disks.  At the highest redshifts represented in our sample, the image resolution is 2.4 kpc$^{2}$ per pixel (see discussion in section \ref{sec:redshift}).  

Our goal is to use a complete sample of galaxies in the local universe to derive global properties of XUV-disk galaxies based on an estimate of the frequency of such galaxies in our sample.  Our classification scheme is sufficient to meet these goals.  In contrast to the classification scheme of T07, which distinguishes XUV-disks based on the morphology of the extended star formation and thus is potentially linked to various formation scenarios, ours consists of one umbrella term for all galaxies with star formation beyond the previously expected radial limits.  This system does not allow us to characterize different types of XUV-disks, but it does allow us to identify XUV-disks at moderate redshifts.  Indeed, it is only with such a volume-limited sample extending to moderate redshifts that we can attempt to estimate the global properties of galaxies with XUV-disks.  This work also serves as a way of determining the feasibility of detecting XUV-disks at larger distances and assessing the incompleteness of such detections.

\subsection{Examples}
\label{sec:examples}

\begin{figure}
\epsscale{1.2}
\plotone{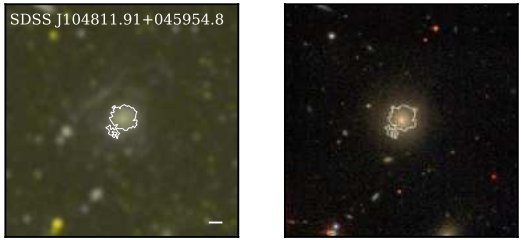}
\vspace{-0.3in}
\caption{\emph{GALEX} and SDSS color imagery of an XUV-disk galaxy.  On the left is the \emph{GALEX} image of a 4$\arcmin$ by 4$\arcmin$ field of view centered on the galaxy (FUV is blue; NUV is yellow).  The white contour indicates the position at which the FUV surface brightness is $\mu_{FUV}$=27.25 ABmag arcsec$^{-2}$.  The white scale bar represents 10 kpc.  On the right is the SDSS DR7 \emph{gri} color composite image of the same field of view.} 
\label{fig:168} 
\end{figure}

Figures 1-3 show a representative sample of galaxies we identified as XUV-disk galaxies, chosen to showcase the diversity of galaxies with extended star formation.  In each figure the 4$\arcmin$ $\times$ 4$\arcmin$ \emph{GALEX} color composite image centered on the galaxy in question is shown on the left.  The contour corresponding to $\mu_{FUV}$=27.25 ABmag arcsec$^{-2}$ is indicated in white.  On the right is the SDSS \emph{gri} color composite image on the same scale.

The optical image of the object in Fig. \ref{fig:168} shows a red (NUV-\emph{r} = 4.89) centrally-concentrated (R$_{90}$/R$_{50}$ = 3.0) galaxy (characteristic of a morphologically early-type galaxy) with a very bright center whose surface brightness falls off quickly.  The \emph{GALEX} image on the left shows that the UV flux drops off much more rapidly than the optical light.  Beyond the optical extent of the object is at least one prominent spiral arm as well as a number of bright UV complexes that constitute the XUV emission.  The XUV emission around this galaxy is similar to the XUV-bright complexes of prototypical XUV-disks \citep{GilDePaz05, Thilker05}; this is one of the more obvious cases of an XUV-disk.

\begin{figure}
\epsscale{1.2}
\plotone{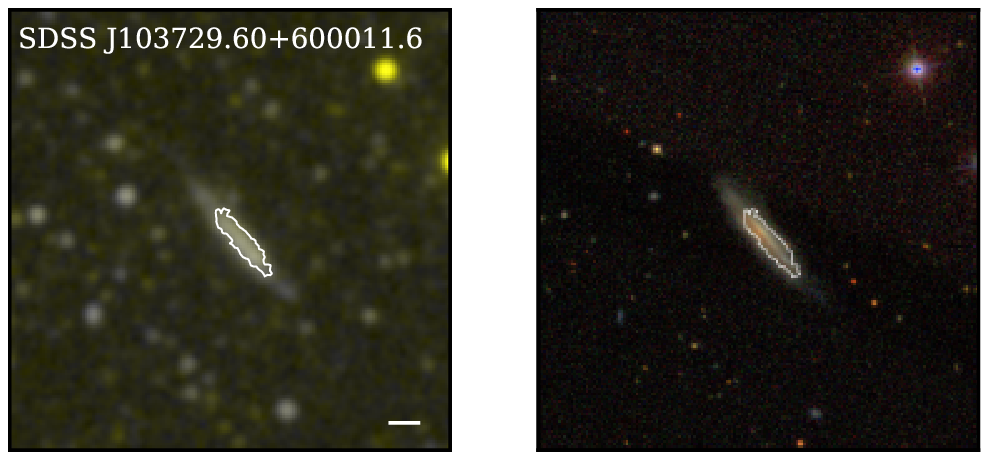}
\vspace{-0.3in}
\caption{\emph{GALEX} and SDSS color imagery of an XUV-disk galaxy.  See caption of Fig. \ref{fig:168} for an explanation of images.} 
\label{fig:831}  
\end{figure}

Fig. \ref{fig:831} shows an edge-on disk galaxy with color NUV-\emph{r} = 3.89 and a prominent red center presumably due to dust extinction.  The faint UV emission apparent at both ends of the galaxy in the \emph{GALEX} image is indicative of a region of XUV emission beyond the expected extent of star formation in the disk.  This galaxy also appears to have a slight warp in its XUV-disk.  Unlike other samples studied for the presence of XUV-disks, our sample is blind to inclination angle.  Because the critical isophote we use is not dust-corrected, our method is less effective for edge-on galaxies because the precise location of the star formation threshold is not obvious.  On the other hand, projection effects observed in edge-on galaxies might make the XUV emission more apparent than it would be in a face-on galaxy.

\begin{figure}[t]
\epsscale{1.2}
\plotone{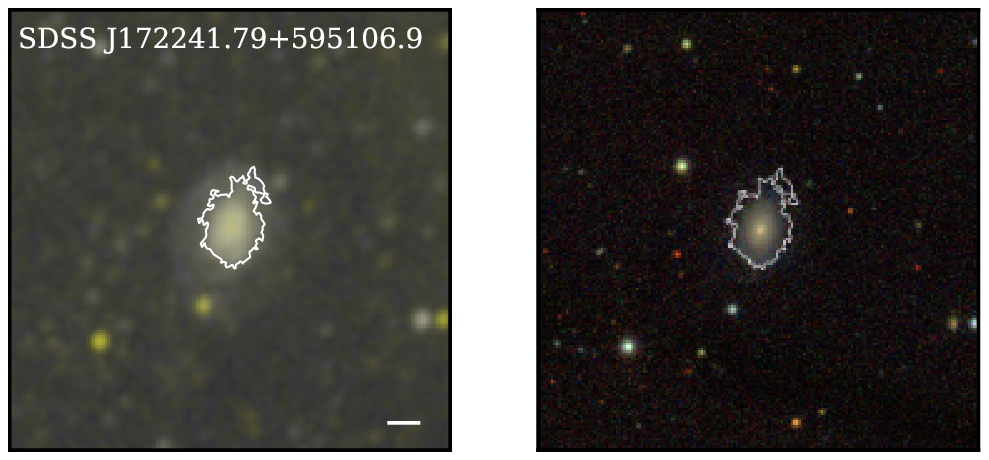}
\caption{\emph{GALEX} and SDSS color imagery of an XUV-disk galaxy.  See caption of Fig. \ref{fig:168} for an explanation of images.} 
\label{fig:37} 
\end{figure}

Fig. \ref{fig:37} shows a blue (NUV-\emph{r} = 1.14) disk + bulge galaxy with R$_{90}$/R$_{50}$ = 2.8.  Within the contour defining the star formation threshold is bright UV emission that mirrors the optical image.  Outside of the contour is an expansive LSB zone of UV emission.  

\begin{figure}
\begin{center}
\vspace{-0.3in}
\epsscale{1.2}
\plotone{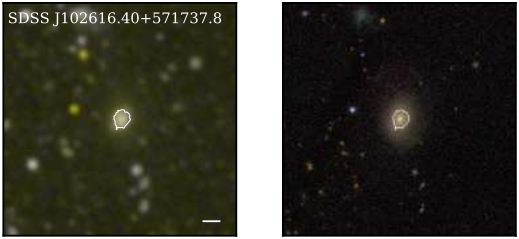}
\caption{\emph{GALEX} and SDSS color imagery of an XUV-ambiguous galaxy. See caption of Fig. \ref{fig:168} for an explanation of images.} 
\label{fig:252} 
\end{center}
\end{figure}

In Fig. \ref{fig:252} we present an example of an XUV-ambiguous galaxy.  This red (NUV-\emph{r} = 5.5) early-type (R$_{90}$/R$_{50}$=3.3) galaxy has a faint ring-like feature in the optical image which could be suggestive of an XUV-disk.  In the \emph{GALEX} image we see UV flux extending just beyond the star formation contour in all directions.  Although this could be indicative of an XUV-disk, we classified this galaxy as XUV-ambiguous because the symmetry of the UV emission suggests that it might be a product of the \emph{GALEX} PSF.  We have not definitively determined that the \emph{GALEX} PSF is the issue here because we have not deconvolved the images.  We also note here that the XUV emission associated with definite XUV-disks generally extends further beyond the star formation threshold than XUV emission around what we consider to be XUV-ambiguous galaxies. 

\section{RESULTS AND DATA ANALYSIS}
\label{sec:results}

\tabletypesize{\footnotesize}

\begin{deluxetable*}{lcccc} 
\tablecolumns{5}
\tablecaption{Properties of XUV-disks and XUV-ambiguous Galaxies}
\tablehead{
\colhead{SDSS ID} & \colhead{FUV-\emph{r}$_{\mathrm{XUV}}$} &  
\colhead{R$_{80}$/R$_{\mathrm{XUV}}$}& 
\colhead{Type\tablenotemark{a}} & \colhead{Reject?\tablenotemark{b}}  \\
\colhead{(1)} & \colhead{(2)} & \colhead{(3)} & \colhead{(4)} & \colhead{(5)} 
}
\startdata

SDSS J120123.26-034100.2	&1.78	&1.7	&D	&N\\
SDSS J171637.43+582442.9	&2.63	&0.9	&D	&N\\
SDSS J171843.70+580806.6	&3.73	&2.0	&A	&N\\
SDSS J171841.08+603629.5	&4.67	&1.4	&D	&N\\
SDSS J172241.79+595106.9	&3.42	&1.0	&D	&N\\
SDSS J140415.84+040643.9	&6.75	&7.3	&A	&Y\\
SDSS J140448.83+045851.7	&7.10	&7.4	&A	&Y\\
SDSS J223619.86+141852.3	&4.67	&4.2	&D	&N\\
SDSS J223649.85+142313.0	&5.13	&6.2	&D	&Y\\
SDSS J103245.73+585137.6	&5.96	&3.6	&D	&Y\\
\enddata
\label{tbl-2}
\tablecomments{This table is available in its entirety in a machine-readable form in the online journal.  A portion is 
shown here for guidance regarding its form and content.}
\tablenotetext{a}{`D' indicates XUV-disk.  `A' indicates XUV-ambiguous disk.}
\tablenotetext{b}{`Y' indicates ``rejects" with FUV-\emph{r}$_{\mathrm{XUV}}$ $>$ 5.0.  `N' indicates galaxies with FUV-\emph{r}$_{\mathrm{XUV}}$ $<$ 5.0.}
\end{deluxetable*}

\begin{figure*}[t]
\begin{center}
\epsscale{1.0}
\plotone{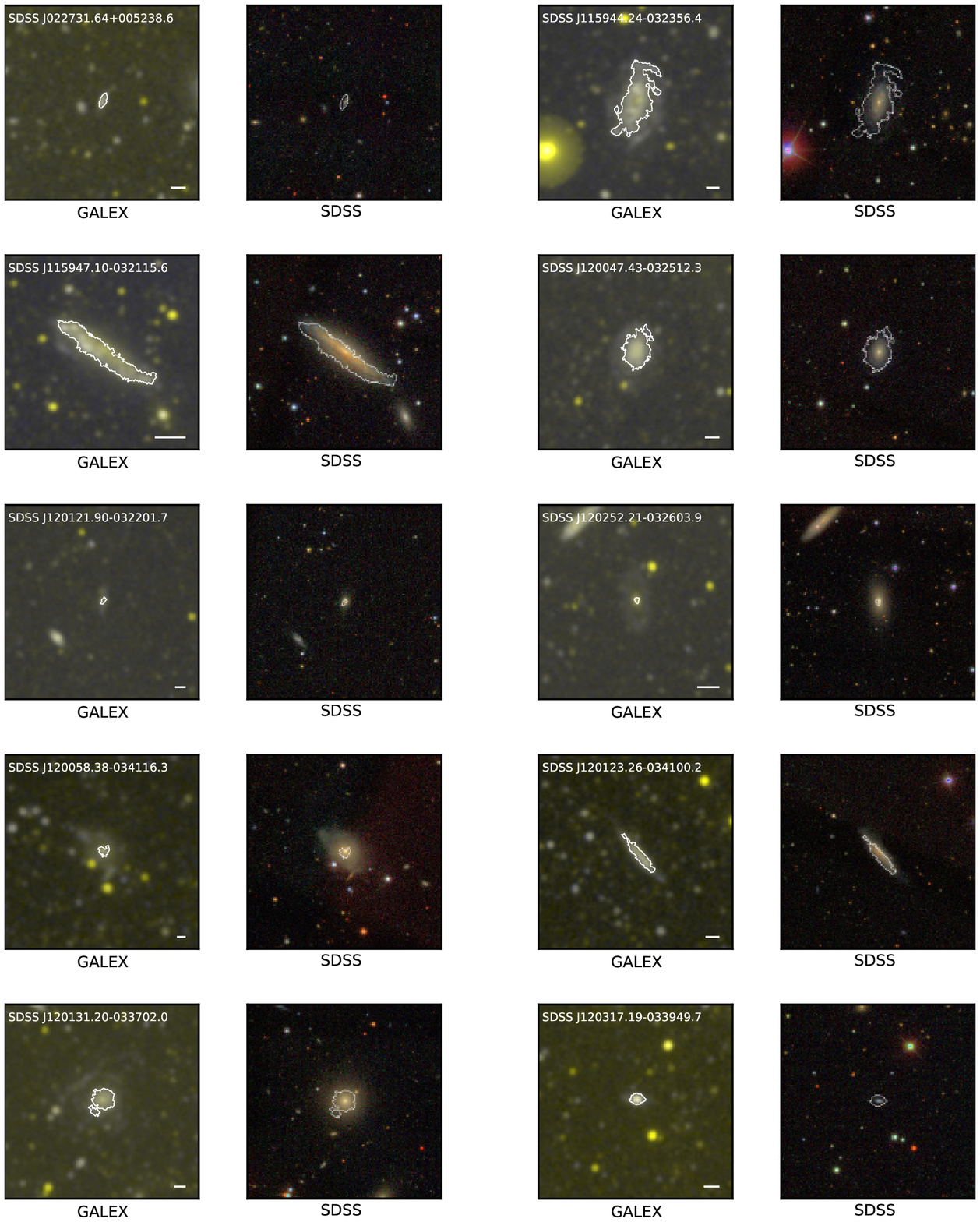}
\caption[]{The 24 XUV-disk galaxies in the sample.  See caption of Fig. \ref{fig:168} for an explanation of images.}  
\label{fig:mosaic1} 
\end{center}
\end{figure*} 

\begin{figure*}[t]
\epsscale{1.0}
\plotone{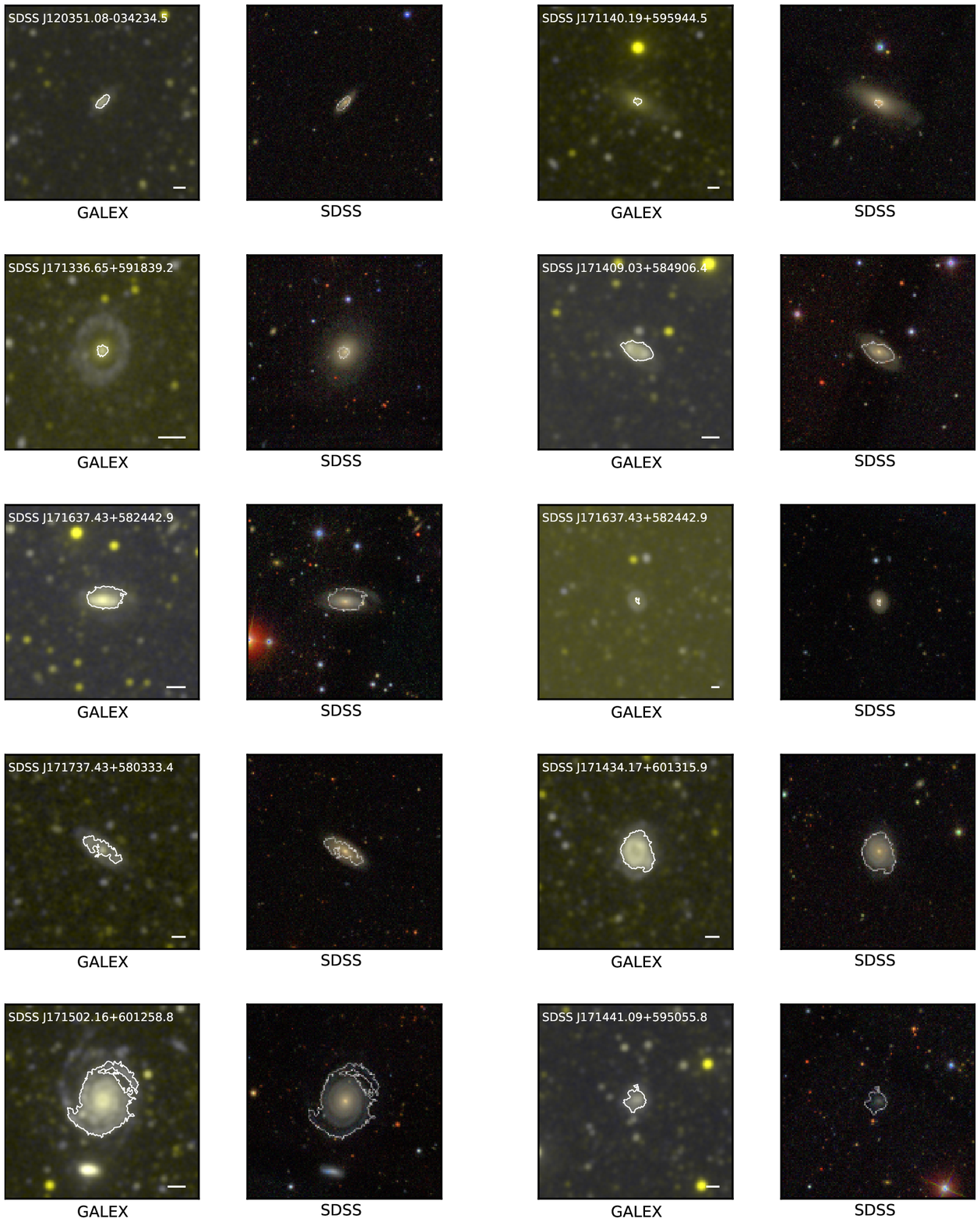}
\caption[]{\emph{Continued}}
\label{fig:mosaic2} 
\end{figure*} 

\begin{figure*}[t]
\epsscale{1.0}
\plotone{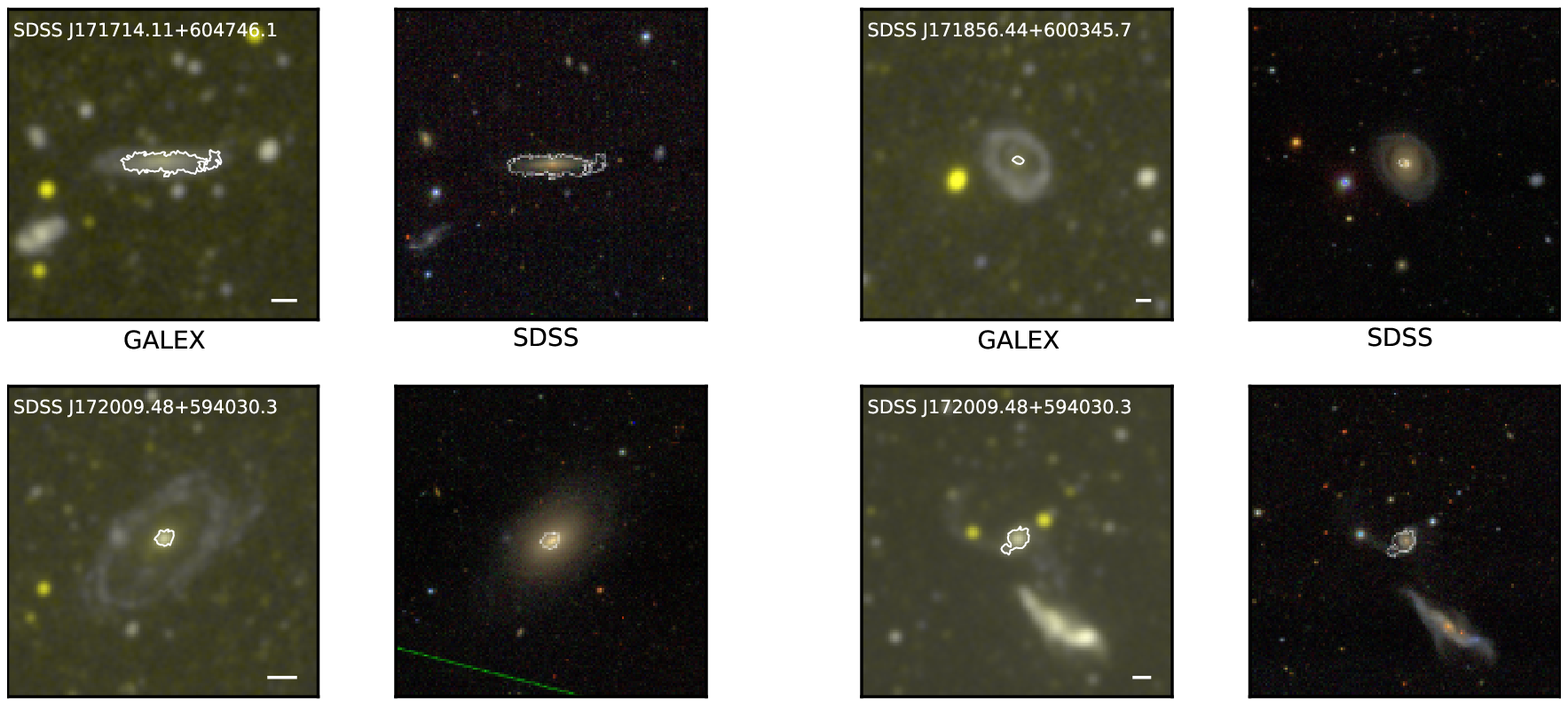}
\caption[]{\emph{Continued}}
\label{fig:mosaic3} 
\end{figure*} 

We identified 24 galaxies out of our sample of 561 as XUV-disk galaxies; that is, more than 4\% of our sample shows prominent UV emission in the outer extents of the galaxy.   We identified another 56 galaxies as XUV-ambiguous.  Thus, we report a total XUV frequency of up to 14\%.  The first frequency is a hard lower limit because of redshift effects that lessen our ability to identify XUV-disks at higher redshifts (see section \ref{sec:redshift}) and because of the non-inclusion of XUV-ambiguous galaxies.  See Fig. \ref{fig:mosaic1} for UV and optical images of XUV-disks and Table 2 for properties of all XUV-disks and XUV-ambiguous galaxies.

As the examples in the previous section demonstrate, XUV-disks are a heterogeneous group of galaxies.  There are however, some features that are common to a number of galaxies that exhibit below-threshold star formation in their outer extents.  Among these are rings and spirals arms beyond the central star-forming region of a galaxy.  In our sample 4 XUV-disk galaxies have UV rings and 6 have spiral arms or fragments of spiral arms beyond the expected star formation threshold.  One XUV-disk galaxy is clearly undergoing an interaction.  Such features could be useful in identifying other XUV-disks and determining the origin of XUV-disks. 

\subsection{Redshift Distribution of XUV-disk Galaxies}
\label{sec:redshift}

\begin{figure}
\epsscale{1.2}
\plotone{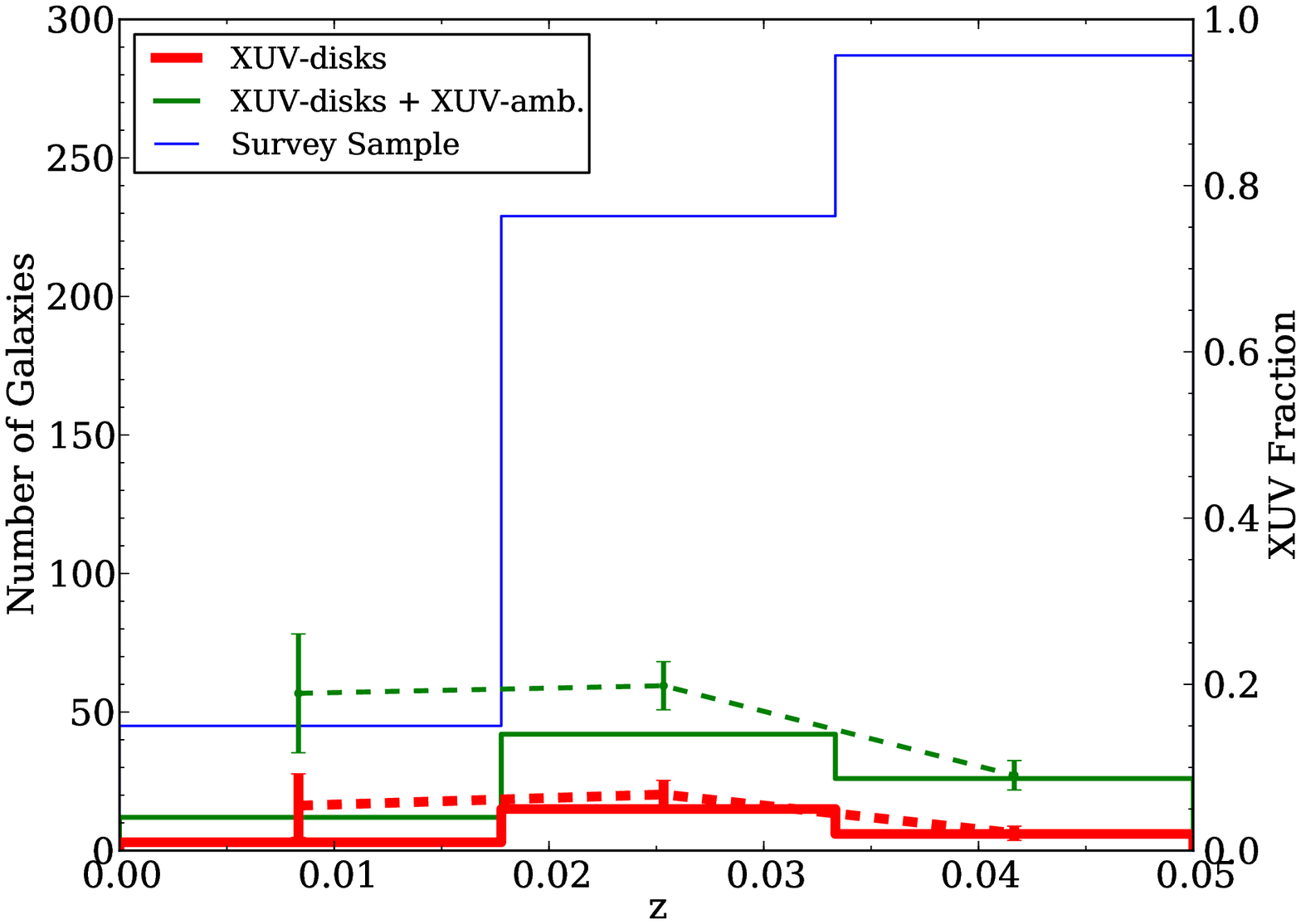}
\caption{The number and fraction of XUV-disk galaxies as a function of redshift.  The thin blue histogram shows the distribution of the entire sample; the thick red histogram shows only the XUV-disk galaxies; the medium green histogram shows both XUV-disk galaxies and XUV-ambiguous galaxies.  The thick red and medium green dashed lines show the fraction of galaxies with XUV emission for XUV-disks only and for XUV-disks in addition to XUV-ambiguous galaxies, respectively.} 
\label{fig:redshift} 
\end{figure}

Fig. \ref{fig:redshift} shows the distribution of XUV-disk galaxies in terms of redshift. Our volume-limited sample is heavily weighted towards objects with $z >$ 0.02.  Not surprisingly, the absolute number of XUV-disk galaxies in our sample is also weighted towards objects with $z >$ 0.02, but the proportion of sample galaxies with XUV-disks is higher at low redshifts.  The data for our lowest and middle redshift bins show that almost 7\% of galaxies at $z<0.01$ have detected XUV-disks.  The fraction of XUV-disk galaxies drops to less than 3\% for galaxies with $0.03<z<0.05$.  This trend with redshift is even more pronounced when we include XUV-ambiguous cases.  Note that when both XUV-disks and XUV-ambiguous galaxies are included, the XUV fraction at low redshifts approaches 20\%; taking into account the size of the error bars, this almost matches the frequency of XUV-disks, 29\%, found in T07's sample.  That the XUV fraction drops with redshift is an indication of the incompleteness of our XUV-disk sample beyond z=0.025 (d $\sim$ 107 Mpc).  In the following analysis we do not adjust our results to account for this dependency on redshift; in our discussion we emphasize that all numbers should be considered lower limits.  In Section \ref{sec:redshift_correction} we explore how a redshift correction would change our results.

\begin{figure}
\epsscale{1.2}
\plotone{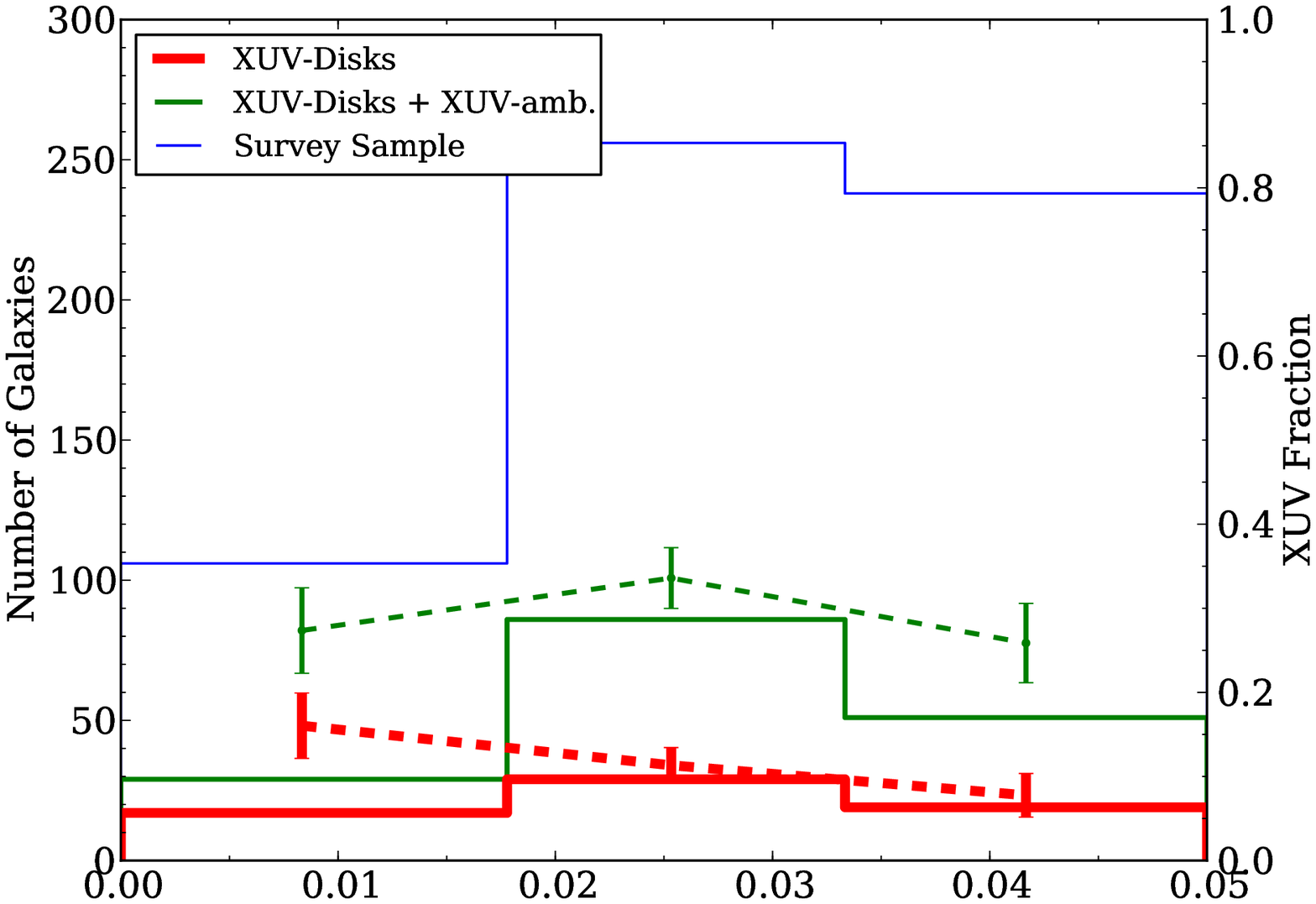}
\caption{Same as Fig. \ref{fig:redshift} but with test galaxies generated with GALFIT.} 
\label{fig:galfit} 
\end{figure}

Almost certainly, this trend with redshift is not a real trend but simply a result of the fact that the classification was done visually and our ability to visually classify galaxies is dependent on the spatial resolution of the images and thus the redshift of each galaxy.  In order to test the conjecture that the observed XUV-disk redshift distribution is due to a selection effect, we used GALFIT \citep{Peng02} to generate a test sample of \emph{GALEX} FUV-band images of artificial galaxies.  Each test galaxy is the sum of an inner disk component and an outer spiral disk component with varying parameters.  Each component has an apparent FUV magnitude between 16.0 and 18.0 at z=0.01.  The inner disk has a half-light radius ranging from 3.5 to 17.5 kpc; the outer disk has a half-light radius ranging from 10.5 to 52.5 kpc.  The Sersic index for the inner disk is set to 1.0 while the Sersic index for the outer disk component ranges from 0.1 to 0.5.  All test galaxies are face-on.  We applied the \emph{GALEX} PSF and noise to each image to mimic the actual images used for the survey.  We then placed each test galaxy at five redshifts from z=0.01 to z=0.05 and used the methodology described above to identify the XUV-disk galaxies in a random selection of the test sample.  Fig. \ref{fig:galfit} shows that we recovered the redshift trend reported above, supporting the hypothesis that the redshift trend of XUV-disk galaxies is not real but a function of the image resolution.  Despite the incompleteness with respect to redshift, our results appear to show that deep \emph{GALEX} imaging can detect a significant number of XUV-disks out to z=0.025 (d $\sim$ 107 Mpc).  We note that this test is limited because such simple models lack the complexity of XUV-disks found in nature.

To further test our ability to identify XUV-disk galaxies at moderate redshifts
we artificially redshifted 35 (out of 54) of the galaxies selected in
T07 to be Type 1 and Type 2 XUV-disks.  The 35 galaxies were chosen because they had readily available NIR photometry.  We used FUV intensity maps of these galaxies compiled as part of the \emph{GALEX} Nearby Galaxies Atlas \citep{GilDePaz07} as a model intensity distribution and then generated
maps for a galaxy of equivalent luminosity at redshifts 0.01, 0.02,
0.03, 0.04 and 0.05.  Simulated intensity maps were generated assuming
a 30000 s total exposure time, with photon contribution from both
model and sky background.  Images were convolved to produce an
appropriate \emph{GALEX} PSF \citep{Morrissey07} and then analyzed using the
same methods described above for identifying XUV-disks.  

\begin{figure}
\epsscale{1.2}
\plotone{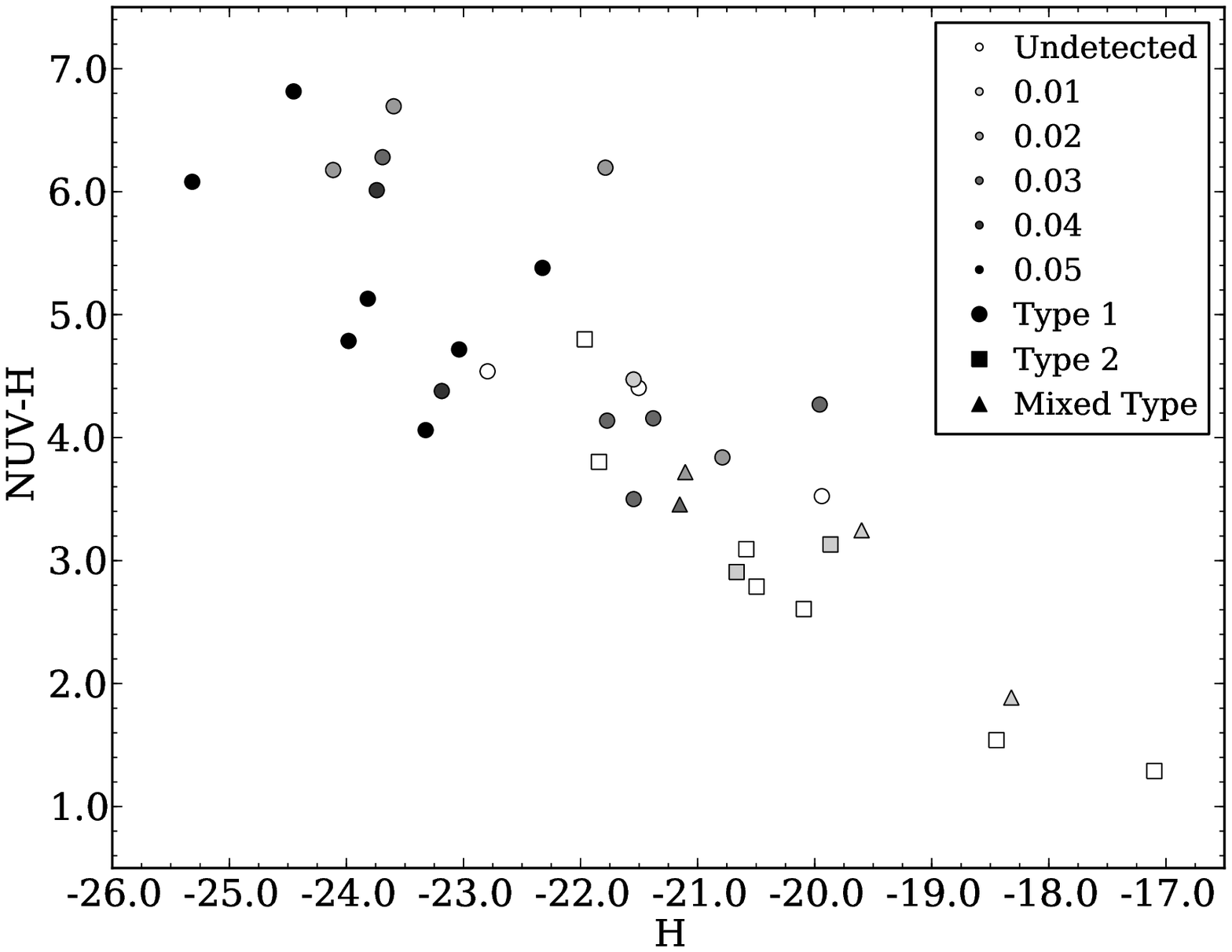}
\caption{The maximum redshift at which an XUV-disk was detected around 35 artificially redshifted galaxies from the sample of \citetalias{Thilker07}.  Galaxies labeled ``undetected" did not exhibit obvious XUV-disks at any of the redshifts we considered.} 
\label{fig:xuvsim} 
\end{figure}

For each galaxy we determined the maximum redshift out to which we
would identify it as an XUV-disk. In Fig. \ref{fig:xuvsim} we show the results of this analysis, with the maximum redshift indicated as a function of
the galaxy's color, NUV-H, and H-band absolute magnitude, as a proxy
for stellar mass.  We find, not surprisingly, that the more luminous
galaxies are easier to detect out to higher redshift.  Additionally
our methods appear to be more effective at identifying Type 1 vs. Type
2 XUV disks, as discussed above.  We found that we could identify XUV-disks out to a median redshift of z=0.02 (z=0.03 for Type 1),
consistent with our observational findings and GALFIT simulated
results discussed above.

There are only 2 (out of 23, or $<$ 9\%) Type 1 XUV-disks in the subset of T07's sample that we studied which would not be considered XUV-disks according to our criteria at redshifts as nearby as 0.01.  However, our classification scheme would allow us to classify the galaxies appropriately as XUV-disks using their actual UV images at their actual distances.  This emphasizes that our classification scheme is generally consistent with that of T07's Type 1 classification scheme but is limited by redshift.

We do not classify as XUV-disks a much higher fraction of T07's Type 2 XUV-disk galaxies.  In fact we only classified 2 out of 9 of T07's Type 2 XUV-disks in the subset that we studied as XUV-disks according to our classification scheme.  This inconsistency is expected because our classification scheme is dependent on UV flux beyond the star formation threshold whereas T07's Type 2 classification criterion is a blue LSB region \emph{within} the FUV star formation threshold.  Although our methodology does recover many of T07's identifications of XUV-disks, we emphasize that our goal is not to reproduce T07's study at larger distances, which distinguished between different morphologies of XUV emission, but to develop an alternative, complementary method for identifying and characterizing extended UV emission around galaxies at moderate redshifts in order to estimate global properties of the population.

\subsection{Morphology of XUV-disk Galaxies}
\label{sec:morphology}

\begin{figure}
\epsscale{1.2}
\plotone{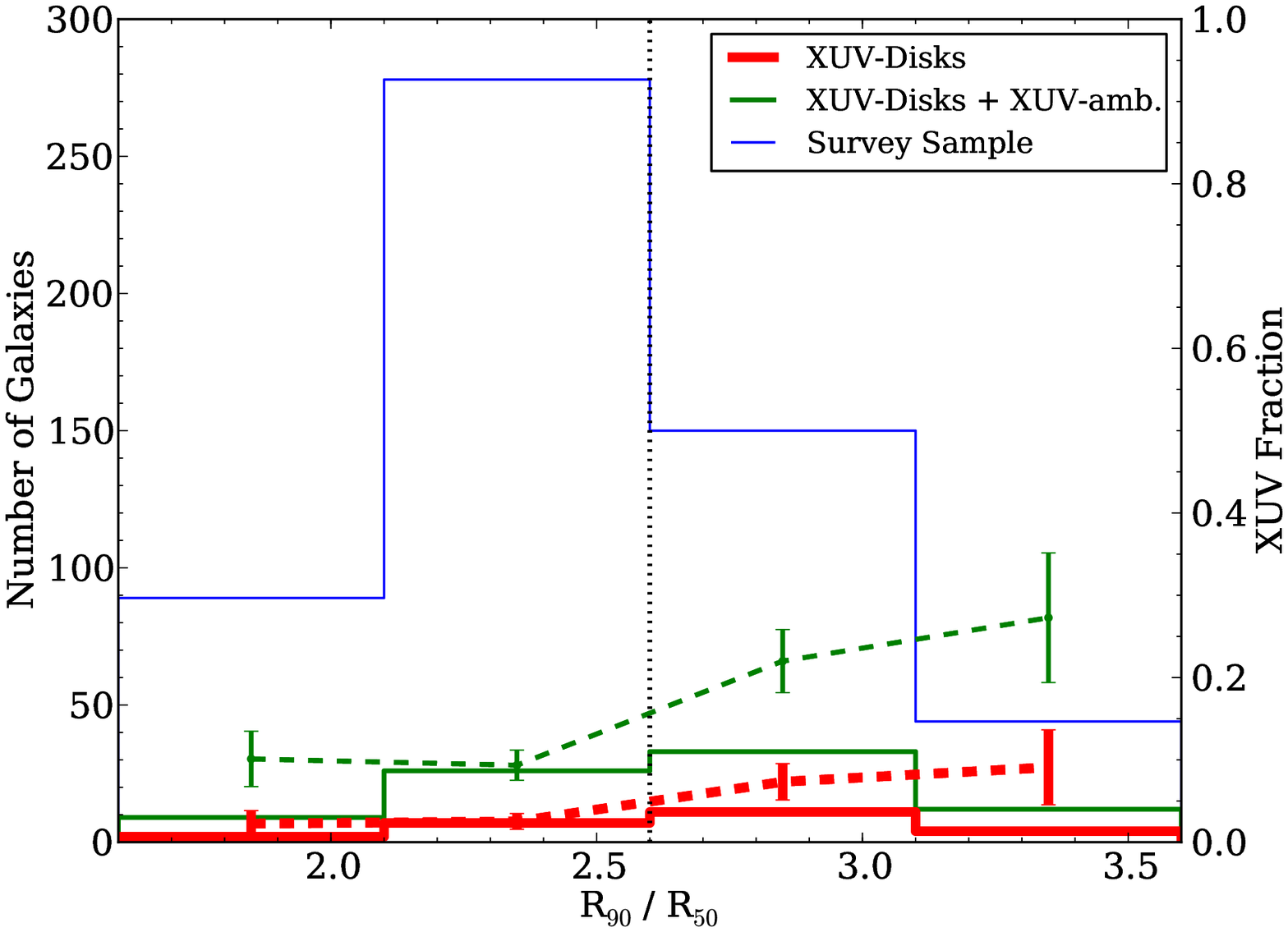}
\caption{The number and fraction of XUV-disks as a function of concentration index (C=R$_{90}$/R$_{50}$).  The dotted vertical line indicates the division between early-type galaxies (C $>$ 2.6) and late-type galaxies (C $<$ 2.6).  The labels are the same as in Fig. \ref{fig:redshift}.} 
\label{fig:concentration} 
\end{figure}

In order to assess the origin and frequency of XUV-disk galaxies, we first investigated the properties of galaxies with and without XUV-disks.  Fig \ref{fig:concentration} shows the distribution of our sample in terms of morphology.  We used the SDSS R$_{90}$ and R$_{50}$ \emph{r}-band values to compute the concentration index C=R$_{90}$/R$_{50}$ and used C as a proxy for the morphology of each galaxy.  According to \citet{Strateva01}, the division between early-type and late-type galaxies occurs at C=2.6; objects with C $>$ 2.6 are considered early-type galaxies and objects with C $<$ 2.6 are considered late-type galaxies.  Our full sample is slightly biased towards late-type objects, median C=2.44.  The population of XUV-disk galaxies in our sample has a median concentration (C=2.68) which is suggestive of more-evolved galaxies.  The median does not change if we include XUV-ambiguous galaxies.  Note that this figure shows that the fraction of XUV-disks increases with concentration index.  This trend suggests that the XUV-disk phenomenon is more common among early-type galaxies with a high concentration index.  An analysis of the colors and masses of the galaxies in our sample provides further evidence that this is the case.

\subsection{Colors and Stellar Masses of XUV-disk Galaxies}
\label{sec:colors}

\begin{figure}
\epsscale{1.2}
\plotone{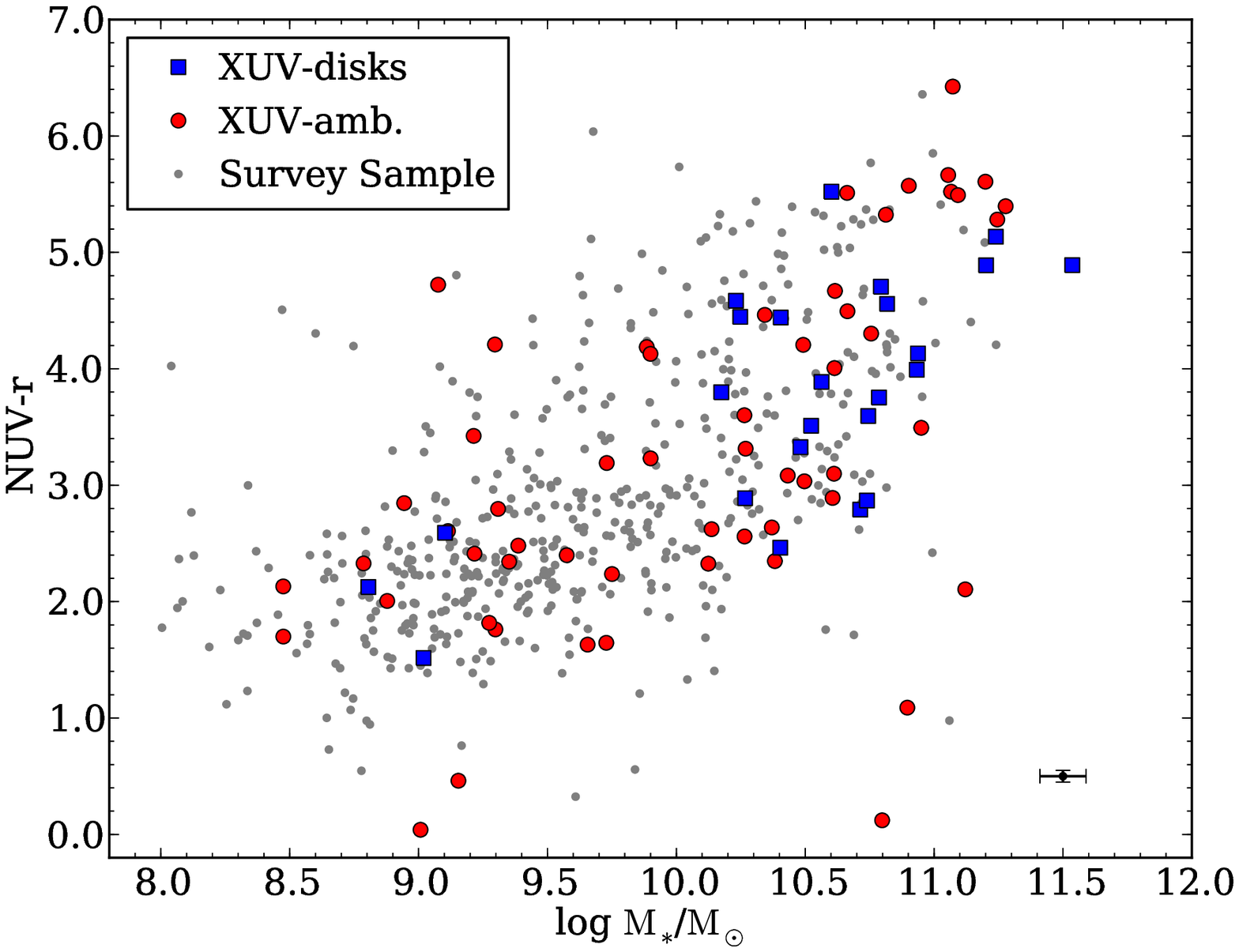}
\caption{NUV-\emph{r} vs. log $M_*$/$M_{\odot}$ for the sample.  XUV-disks are represented by blue closed squares, XUV-ambiguous galaxies by red closed circles, and galaxies with no XUV emission by gray dots.}
\label{fig:mstar} 
\end{figure}

Fig \ref{fig:mstar} shows NUV-\emph{r} vs. log $M_*$/$M_{\odot}$ for the sample and confirms that a large population of XUV-disk galaxies are massive red early-types.  A first glance at the plot shows that XUV-disks and XUV-ambiguous galaxies (represented by closed blue squares and closed red circles, respectively) are scattered fairly evenly throughout the plot.  A closer look shows that a much higher fraction of the red, massive galaxies (NUV-\emph{r} $>$ 3 and log $M_*$/$M_{\odot}$ $>$ 10.0) in the sample are identified as either XUV-disk galaxies or XUV-ambiguous galaxies.  This finding strongly suggests that XUV-disks are more likely to be detected around early-type galaxies.

\begin{figure}[t]
\epsscale{1.2}
\plotone{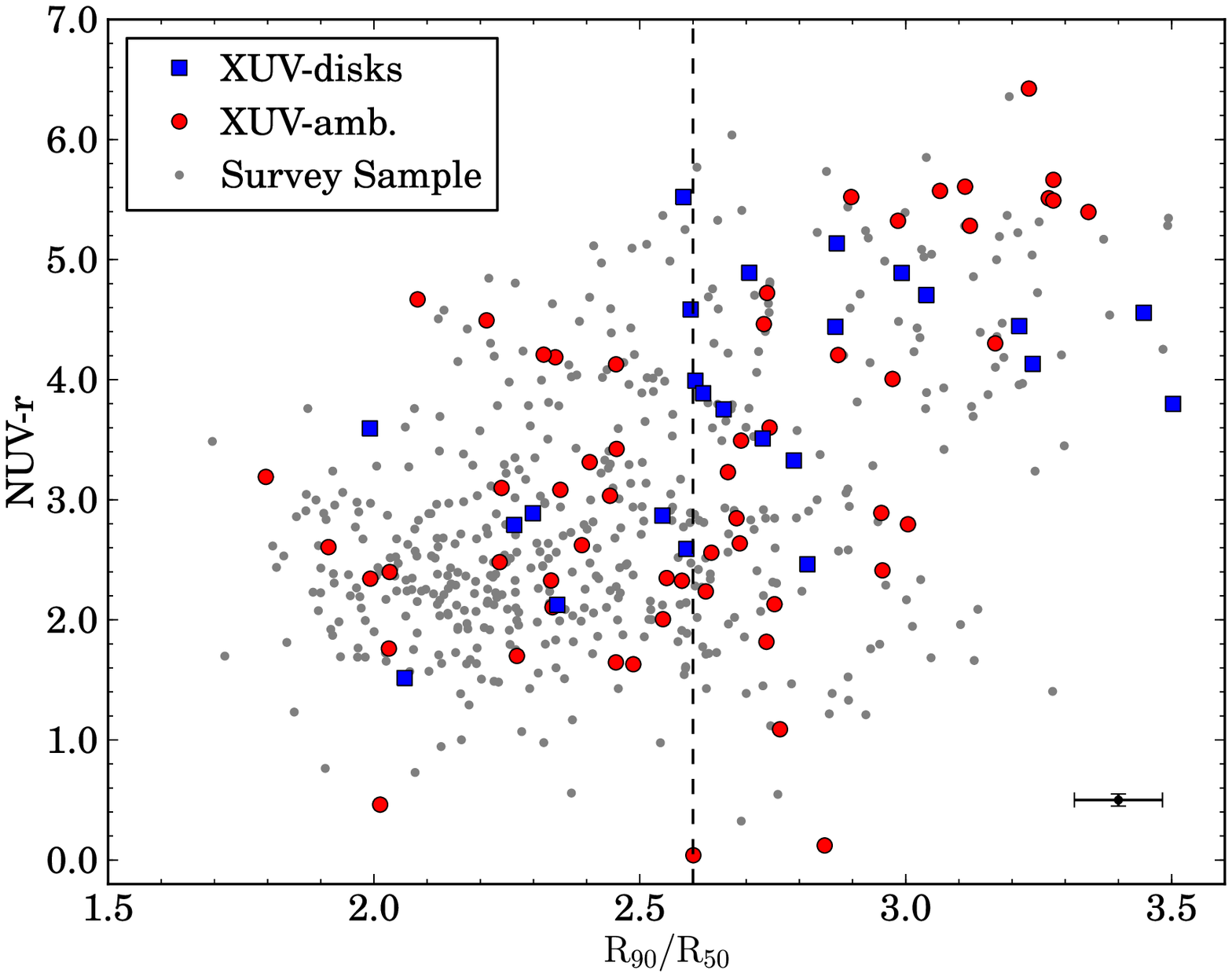}
\caption{NUV-\emph{r} color plotted against concentration index (C=R$_{90}$/R$_{50}$).  The dashed vertical line indicates the division between early-type galaxies (C $>$ 2.6) and late-type galaxies (C $<$ 2.6).  Labels are the same as in Fig. \ref{fig:mstar}.} 
\label{fig:color_c} 
\end{figure} 

The recent finding of \citet{Kannappan09} of a population of blue E/S0 galaxies led us to investigate whether the early-type galaxies we classified as XUV-disks are members of this population.  Fig. \ref{fig:color_c} shows that most of the early-type XUV-disk galaxies have red colors with NUV-\emph{r} $>$ 3 and thus are too red to be members of this category of galaxies.  There are, however, a small number of XUV-disks and XUV-ambiguous galaxies with C $>$ 2.6 and NUV-\emph{r} $<$ 3 that occupy a region of Fig. \ref{fig:color_c} distinct from the regions occupied by blue late-types and red early-types.  Thus, there is some evidence that a fraction of our XUV-classified galaxies may be part of this new population of early-type galaxies whose properties (e.g. color, visual evidence of star formation) are more akin to those of late-type galaxies.  

\subsection{The XUV Emission}
\label{sec:XUV Emission}

\subsubsection{Definition of the XUV Region}
\label{sec:XUVregion}

In order to characterize the extended star formation around XUV-disk galaxies, we computed the FUV and \emph{r}-band magnitudes within the XUV region.  We defined the XUV region as the region between the expected star formation threshold at which $\mu_{FUV}$=27.25 ABmag arcsec$^{-2}$ and the contour at which $\mu_{FUV}$ drops to 29.0 ABmag arcsec$^{-2}$.  The outer limit of the XUV region was chosen to maximize the amount of XUV flux contained within the contour and to minimize the number of neighboring sources that mistakenly fall within the XUV region.  In many cases, the XUV region does not completely encompass each UV complex or the entirety of the UV-bright region that led us to visually classify the galaxy as an XUV-disk or XUV-ambiguous galaxy.  But a rigorous definition of the XUV region allows us to comment further on the characteristics of the XUV emission.  We obtained photometry of the XUV region for all galaxies in the sample to allow us to compare these regions of the galaxies and perhaps clarify our visual classification scheme.  We note that a fraction of the galaxies with no XUV emission do not have XUV regions because they have peak FUV surface brightnesses fainter than 27.25 ABmag arcsec$^{-2}$.

\subsubsection{Color of the XUV Region}
\label{sec:color cut}

\begin{figure}
\epsscale{1.2}
\plotone{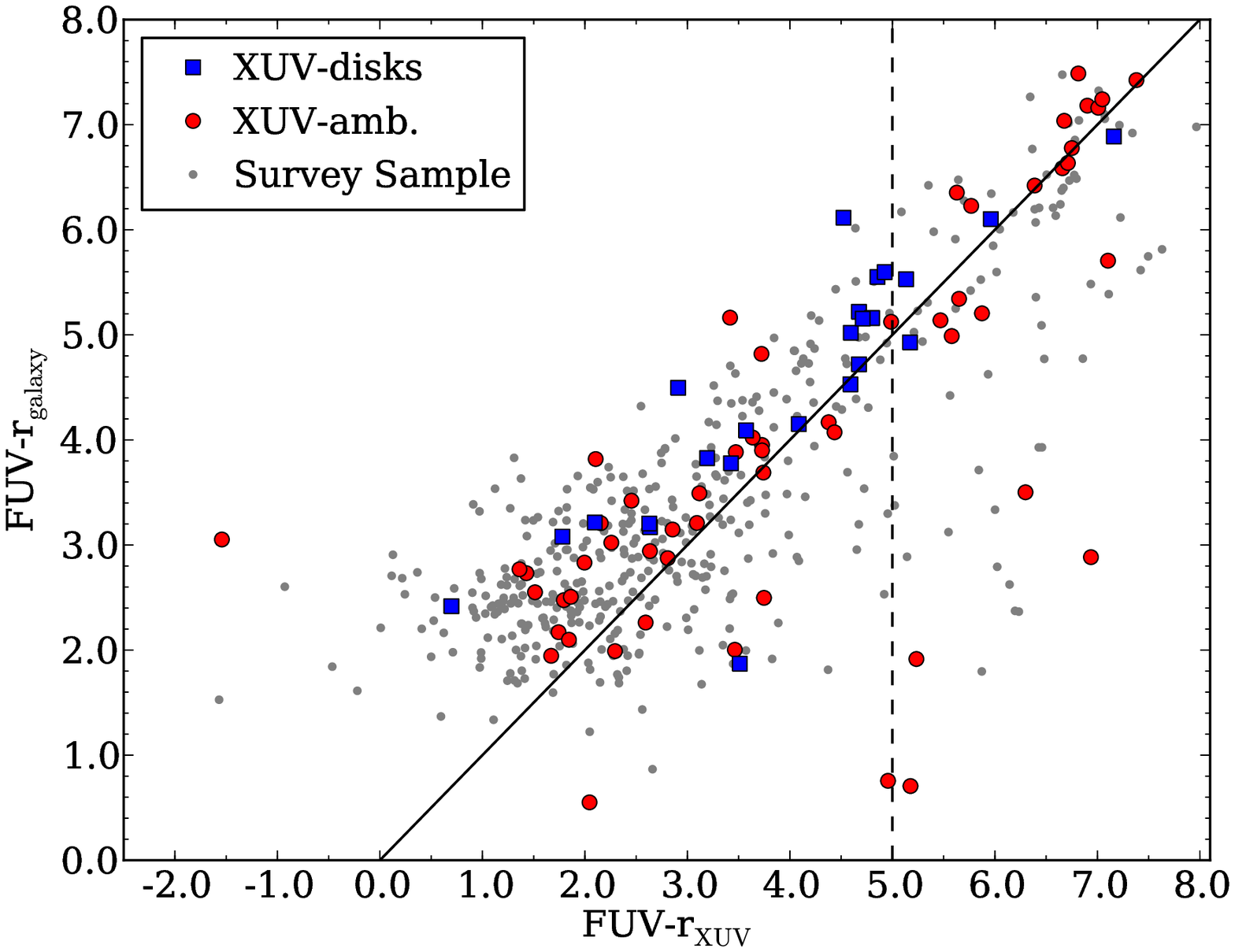}
\caption{FUV-\emph{r} color of each galaxy (FUV-\emph{r}$_{\mathrm{galaxy}}$) plotted against FUV-\emph{r} color of the XUV region (FUV-\emph{r}$_{\mathrm{XUV}}$).  The vertical dashed line shows where FUV-\emph{r}$_{\mathrm{XUV}}$ = 5.0.  The solid line has a slope of 1.  Labels are the same as in Fig. \ref{fig:mstar}.} 
\label{fig:colorcut} 
\end{figure} 

Fig. \ref{fig:colorcut} shows the global FUV-\emph{r} color of each galaxy plotted against the FUV-\emph{r} color of each galaxy's XUV region (FUV-\emph{r}$_{\mathrm{XUV}}$).  Our goal in picking out galaxies with extended UV emission was to characterize the types of galaxies with recent star formation beyond the expected limits; a color cut based on FUV-\emph{r} distinguishes between those XUV-disk galaxies with recent star formation in their XUV regions and those XUV-disk galaxies that contain more evolved stellar populations in their XUV regions.  The vertical dashed line in Fig. \ref{fig:colorcut} indicates FUV-\emph{r}$_{\mathrm{XUV}}$ = 5.0.  \citet{Wyder07} show that the blue star-forming sequence of galaxies extends to FUV-\emph{r} $\sim$ 5.0.  Additionally, \citet{Johnson07} show that galaxies with FUV-\emph{r} $\sim$ 5.0 have a 4000 \AA\ break strength (D$_{n}$(4000); a standard stellar age index) of 1.7.  Assuming an instantaneous burst of star formation, D$_{n}$(4000) = 1.7 corresponds to star formation within the past 1.5 Gyr \citep{Kauffmann03}.  Thus, removing from consideration galaxies with FUV-\emph{r}$_{\mathrm{XUV}}$ $>$ 5.0 restricts our sample to galaxies whose XUV regions had some star formation at most 1.5 Gyr ago.

If we were to include a color cut in our classification scheme, how would our sample of XUV-disks change?  Clearly, the sample would no longer contain many red, massive early-types.  This raises the question of what are the red, massive early-types that we identified as XUV-disks and XUV-ambiguous galaxies whose XUV regions are not very blue at all.  Although the XUV regions are surprisingly red, if we look at galaxies with continually redder XUV regions we uncover galaxies that are redder than their XUV regions.  That the XUV regions around these galaxies are slightly bluer than the colors of the overall galaxies suggests that star formation in the outer regions has proceeded more recently than would be expected given the global colors which suggest that galaxy-wide star formation has ceased.

\begin{figure*}[t]
\epsscale{1.0}
\plotone{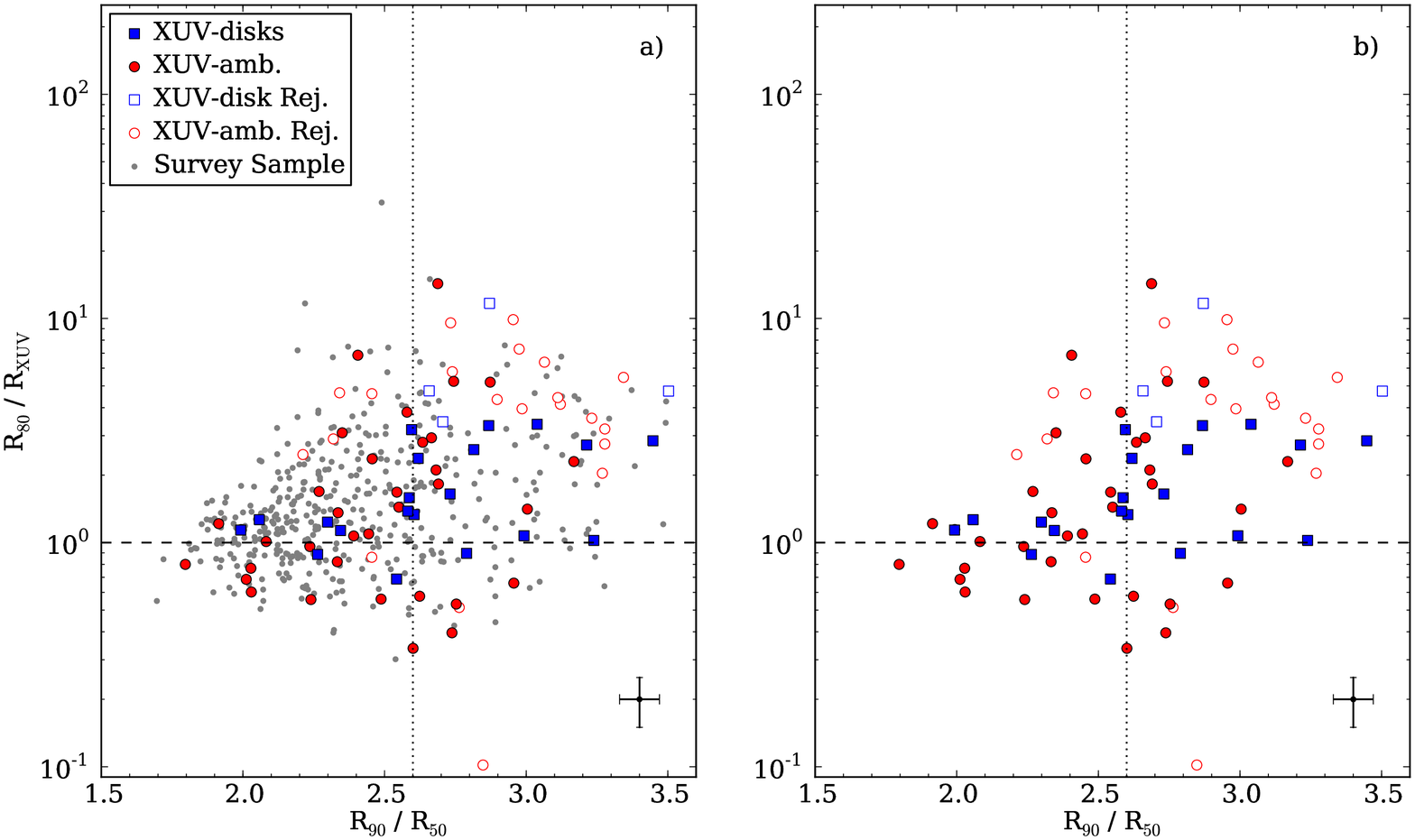}
\caption{R$_{80}$/R$_{\mathrm{XUV}}$ vs. concentration index for all galaxies in the sample (a) and only galaxies with XUV flux (b).  Gray points are sample galaxies with no XUV flux.  Blue closed squares are XUV-disks and red closed circles are XUV-ambiguous galaxies.  Blue open squares are XUV-disk ``rejects" and red open circles are XUV-ambiguous ``rejects."  The dotted vertical line indicates the division between early-type galaxies (C $>$ 2.6) and late-type galaxies (C $<$ 2.6).  The dashed horizontal line shows where R$_{80}$=R$_{\mathrm{XUV}}$.}
\label{fig:moffett} 
\end{figure*} 

In order to assess why a fraction of XUV-disks and XUV-ambiguous galaxies have remarkably red XUV regions, we found the contour that includes 80\% of the r-band light, R$_{80}$.  We define R$_{80}$ as the mean of the projected major and minor axes of this contour. R$_{80}$ should be comparable to K$_{80}$, which defined the inner contour of the LSB region for Type 2 XUV-disks in T07.  In Fig. \ref{fig:moffett} we show R$_{80}$ divided by the extent of the XUV threshold (R$_{\mathrm{XUV}}$; also measured in projection) plotted against the concentration index (C=R$_{90}$/R$_{50}$).  The ratio R$_{80}$/R$_{\mathrm{XUV}}$ and the concentration index both measure how centrally concentrated the star formation is within a galaxy, but the former is directly related to our analysis in that it provides another indicator of the color of the XUV region.  For high R$_{80}$/R$_{\mathrm{XUV}}$,  R$_{80}$ is well beyond the inner FUV threshold so we would expect a substantial amount of \emph{r}-band light to contaminate the XUV region.   

In this figure XUV-disks and XUV-ambiguous galaxies with FUV-\emph{r}$_{\mathrm{XUV}}$ $<$ 5.0 are shown as closed symbols and ``rejects," those with FUV-\emph{r}$_{\mathrm{XUV}}$ $>$ 5.0, are shown as open symbols.  Most of the ``rejects" have high R$_{80}$/R$_{\mathrm{XUV}}$ and most of the galaxies with blue XUV regions have low R$_{80}$/R$_{\mathrm{XUV}}$.  This shows that excluding galaxies with high R$_{80}$/R$_{\mathrm{XUV}}$ is comparable to using a FUV-r cut on the color of the XUV region.  Such galaxies have XUV region colors that may be redder than expected, but they still show evidence of recent and unexplained below-threshold star formation.  For that reason, we do not exclude such galaxies in the following analysis, although we do explore how such a cut based on XUV region color would affect our conclusions in the Appendix.

The left panel of Fig. \ref{fig:moffett} includes the entire sample of galaxies and shows that XUV-disks do not stand apart from the rest of the sample.  This is possibly due to the fact that much of the UV flux seen beyond the threshold in general has a very low surface brightness.  Similarly, T07 found that the star formation beyond the star formation threshold in XUV-disks contributed negligibly to the overall star formation in a galaxy.  Except for a few outliers, the ``rejects" do occupy an extreme region of the plot.  They tend to be highly-concentrated galaxies, and their R$_{80}$/R$_{\mathrm{XUV}}$ ratios show that their UV star formation threshold is well within the main part of the star-forming body, which R$_{80}$ approximates.  Thus it is appropriate to discard them as true XUV-disks since their XUV emission is part of the main stellar disk.  

In extending the XUV-disk classification system of T07 to early-type galaxies, \citet{Moffett09} discovered that the XUV threshold at which $\mu_{FUV}$=27.25 ABmag arcsec$^{-2}$, which defined the outer contour of the LSB region, often fell inside K$_{80}$, making it impossible to characterize these galaxies as Type 2 XUV-disks.  They found that K$_{80}$ was within the star formation threshold for only the UV-bluest E/S0s, which also had the highest SFRs in their sample.  

For those galaxies that did have K$_{80}$/R$_{\mathrm{XUV}}$ $<$ 1, the LSB region between K$_{80}$ and R$_{\mathrm{XUV}}$ was not large enough for the galaxy to be considered a Type 2 XUV-disk according to T07's rigorous definition.  \citet{Moffett09} described such galaxies whose LSB regions were small but blue as ``modified'' Type 2 XUV-disks.  ``Modified" Type 2 XUV-disks may represent a class of galaxies that are fundamentally different from Type 2 XUV-disks.  Although we did not search for Type 2 or ``modified'' Type 2 XUV-disks in our sample, it is likely that they would lie in the region of our plot where R$_{80}$ $<$  R$_{\mathrm{XUV}}$.  Rejects with R$_{80}$/R$_{\mathrm{XUV}}$ $<$ 1.0 may in fact be Type 2 or ``modified'' Type 2 XUV-disks.

Finally, we note that the colors of the XUV region provide support for our bimodal classification scheme in which XUV-disk galaxies and XUV-ambiguous galaxies are classified separately.  Only 16\% of the XUV-disk galaxies have FUV-\emph{r}$_{\mathrm{XUV}}$ $>$ 5.0 but over 37\% of the XUV-ambiguous galaxies have FUV-\emph{r}$_{\mathrm{XUV}}$ $>$ 5.0.  

\begin{figure}
\epsscale{1.2}
\plotone{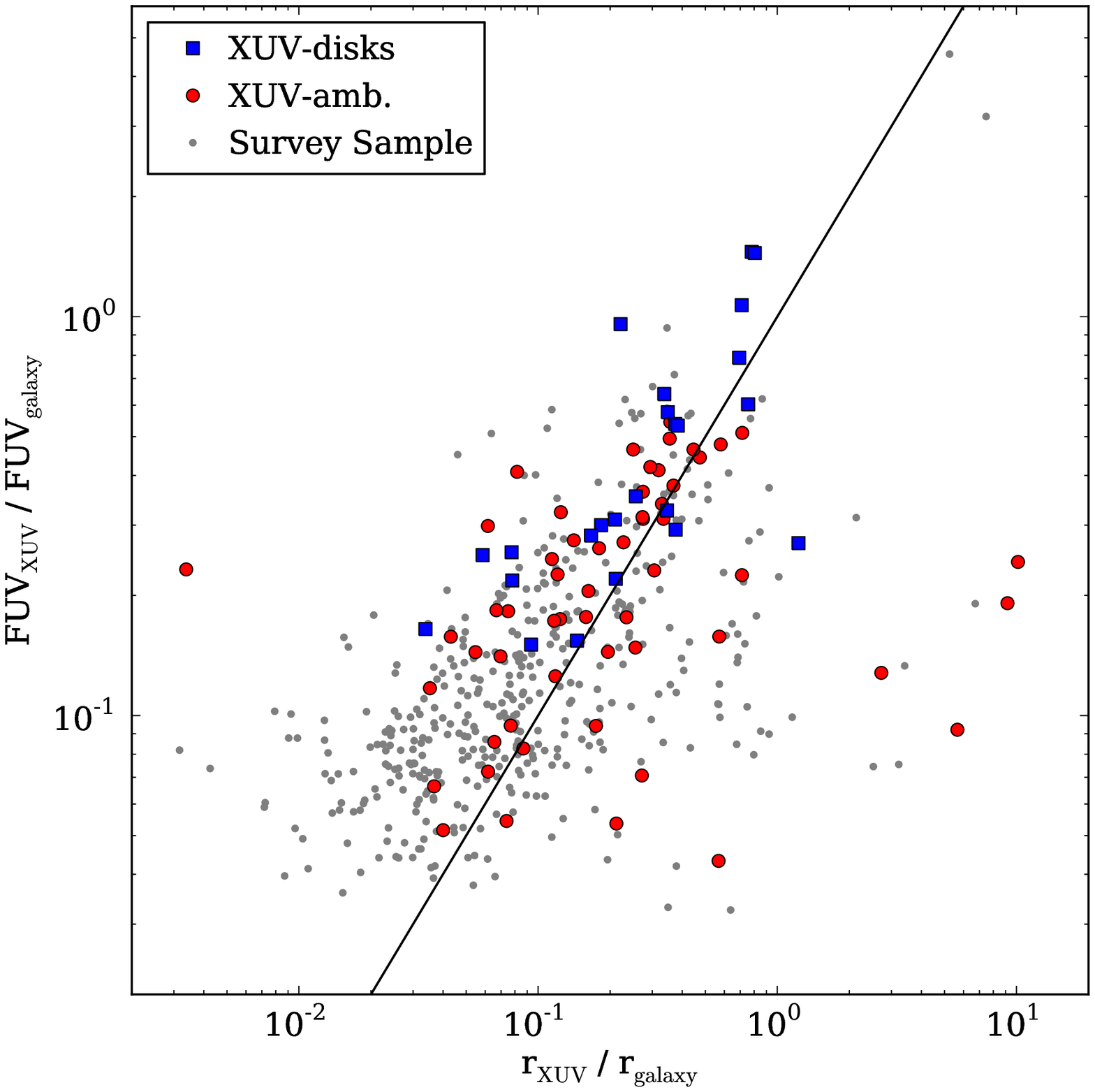}
\caption{The FUV flux in the XUV region compared to the entire galaxy vs. the \emph{r}-band flux in the XUV region compared to the entire galaxy.  The solid line shows where FUV$_{\mathrm{XUV}}$/FUV$_{\mathrm{galaxy}}$=\emph{r}$_{\mathrm{XUV}}$/\emph{r}$_{\mathrm{galaxy}}$.  Labels are the same as in Fig. \ref{fig:mstar}.} 
\label{fig:flux_ratio} 
\end{figure}

Fig. \ref{fig:flux_ratio} presents a variation of Fig. \ref{fig:colorcut} in which we show the FUV flux in the XUV region compared to the global FUV flux (FUV$_{\mathrm{XUV}}$/FUV$_{\mathrm{galaxy}}$) vs. the \emph{r}-band flux in the XUV region compared to the global \emph{r}-band flux (\emph{r}$_{\mathrm{XUV}}$/\emph{r}$_{\mathrm{galaxy}}$).  All XUV-disks appear to have an elevated FUV$_{\mathrm{XUV}}$/FUV$_{\mathrm{galaxy}}$ and no galaxies with very low FUV$_{\mathrm{XUV}}$/FUV$_{\mathrm{galaxy}}$ and \emph{r}$_{\mathrm{XUV}}$/\emph{r}$_{\mathrm{galaxy}}$, suggestive of a lack of star formation in the outer regions, are identified as XUV-disks or XUV-ambiguous galaxies.  

Although the XUV-disks do tend to have high FUV$_{\mathrm{XUV}}$/FUV$_{\mathrm{galaxy}}$, there are many sample galaxies with similar values, and there are a number of XUV-ambiguous galaxies with quite low FUV$_{\mathrm{XUV}}$/FUV$_{\mathrm{galaxy}}$.  The sample galaxies with a high ratio of FUV flux in the XUV region compared to the entire galaxy tend to have UV thresholds that are well within the main body of the galaxy; this artificially enhances the FUV$_{\mathrm{XUV}}$/FUV$_{\mathrm{galaxy}}$ ratio.  XUV-ambiguous galaxies with low ($<$ 10$^{-1}$) ratios tend to have very diffuse, low surface brightness UV flux, which explains why the UV flux in the XUV region is so small compared to the UV flux emitted by the entire galaxy.

Together, Figs. \ref{fig:mstar} - \ref{fig:flux_ratio} show that XUV-disks do not occupy a single region of parameter space and thus may not represent a distinct population of galaxies.  Their diversity, in terms of color, mass, and concentration, along with the diversity of the XUV regions themselves, supports previous suspicions based on the various morphologies of the extended star formation (see Section \ref{sec:examples}) that XUV-disks comprise a heterogeneous population of galaxies.  Thus, galaxies hosting XUV flux are not a class of galaxies unto themselves; rather, they may represent an \emph{evolutionary phase} during which the bulk of star formation occurs beyond the main star forming disk.  We discuss this idea further in Section \ref{sec:disk-building}.

\section{DISCUSSION}
\label{sec:discussion}

\subsection{The Space Density of XUV-disk Galaxies}
\label{sec:frequency}

\begin{figure}
\epsscale{1.2}
\plotone{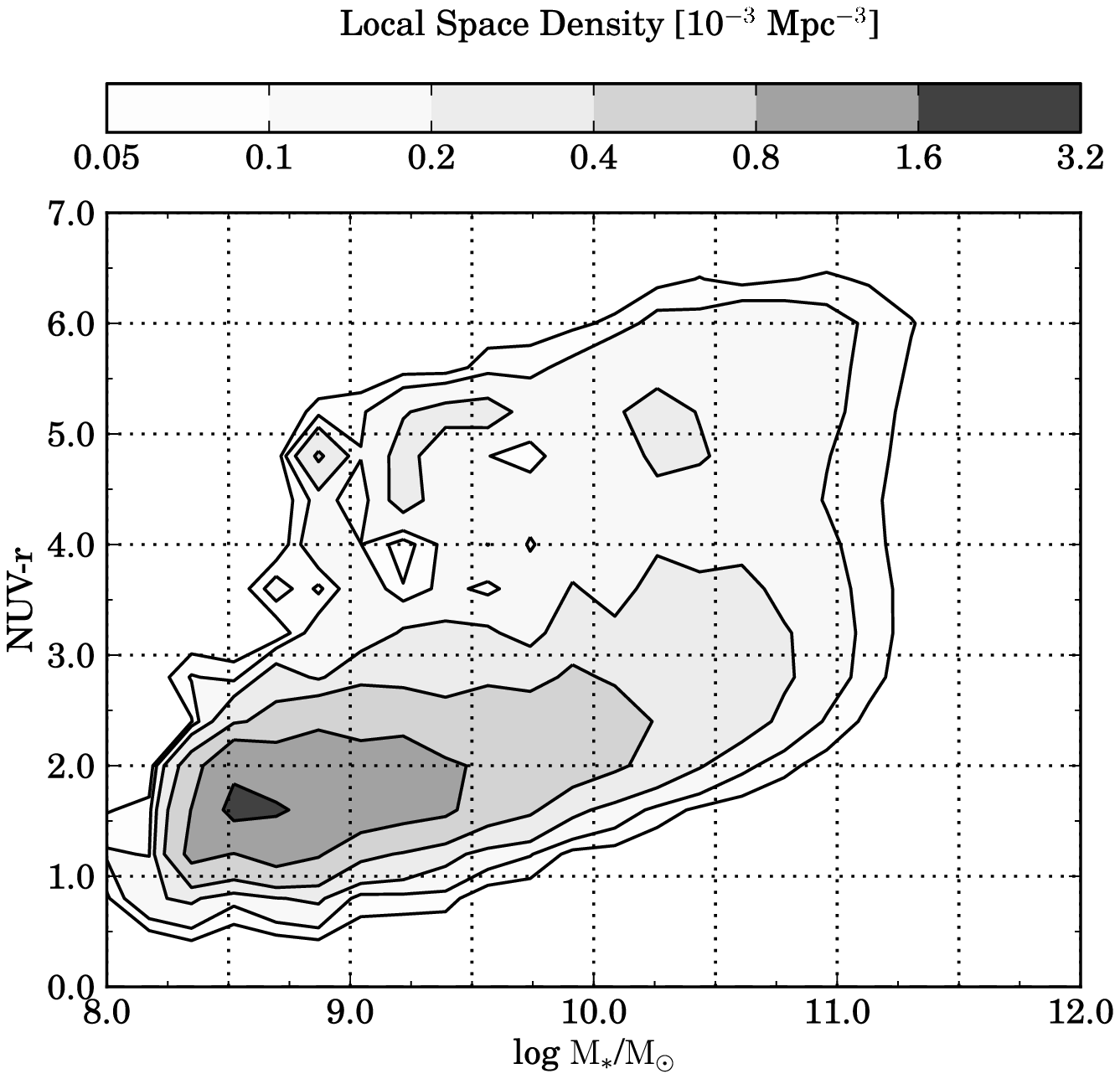}
\caption{The local space density of galaxies, derived from \citet{Wyder07}.  The bins used to determine the space density of XUV-disks are indicated.} 
\label{fig:color_mass_contour} 
\end{figure}

One of the primary goals of this survey was to determine the space density of XUV-disk galaxies in the local universe.  The size of the sample and the unbiased nature of the survey allow us to do this.  In order to compute the space density of XUV-disks we first determined the XUV fraction, or the fraction of galaxies exhibiting XUV-disks in color-stellar mass bins, by volume-averaging the XUV fraction in three evenly-spaced redshift bins.  The space density of XUV-disks is the product of this XUV fraction and the space density (Mpc$^{-3}$) of all galaxies in the local universe, derived from the color-magnitude diagram in \citet{Wyder07}.  We found the space density of galaxies in each color-stellar mass bin by first deriving a color-dependent mass-to-light ratio from the sample in \citet{Wyder07}.  We then used stellar masses from the MPA-JHU catalog as references and a \citet{Chabrier03} initial mass function to convert from absolute magnitude to stellar mass.  The resulting distribution is shown as a contour plot in Fig \ref{fig:color_mass_contour}, with the bins used in the following analysis indicated.  

\begin{figure*}
\epsscale{1.0}
\plotone{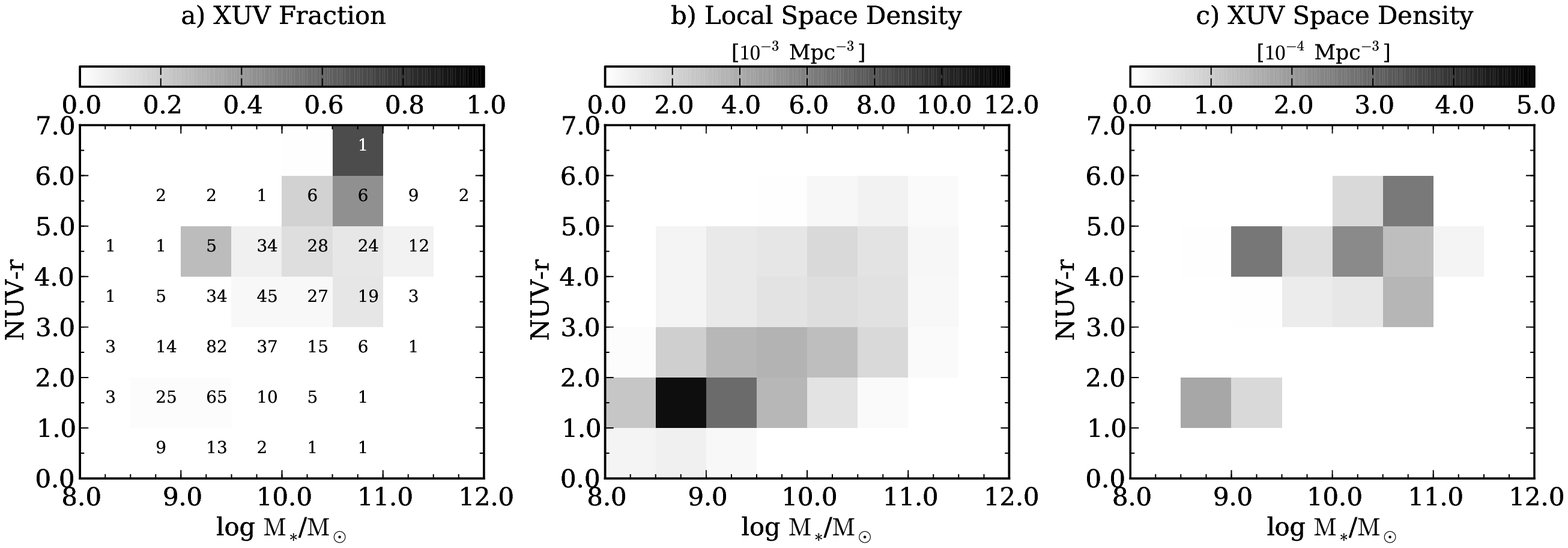}
\caption{The volume-averaged XUV fraction (fraction of galaxies in our sample with XUV-disks), the local space density of galaxies \citep[derived from][]{Wyder07}, and the local space density of XUV-disk galaxies.  The numbers overlaid on the XUV fraction plot indicate the total number of sample galaxies in each bin.  Note that the scales for the plots in the middle and on the right are off by an order of magnitude.} 
\label{fig:xuv_freq_1} 
\end{figure*}

We show in Fig. \ref{fig:xuv_freq_1} (a) the volume-averaged XUV fraction used to find the XUV-disk space density.  The extended peak for galaxies with NUV-\emph{r} $>$ 4 suggests that XUV-disks are more common among transition galaxies in the green valley (3 $<$ NUV-\emph{r} $<$ 5) and red-sequence (NUV-\emph{r} $>$ 5) galaxies.  The local space density of galaxies derived from Fig. \ref{fig:color_mass_contour} is shown in (b).  The product of the XUV fraction and the local space density, in (c), shows that the space density of XUV-disks is fairly even across the sequence.

\begin{figure*}[t]
\epsscale{1.0}
\plotone{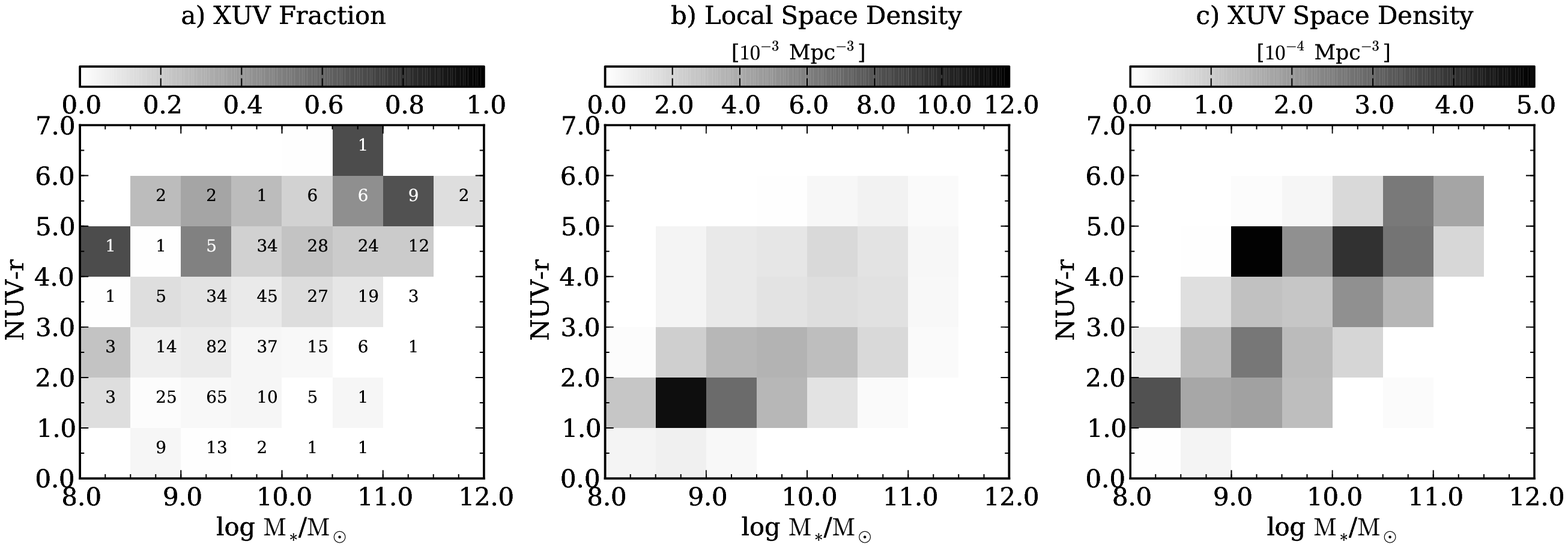}
\caption{Same as Fig. \ref{fig:xuv_freq_1} with XUV-ambiguous galaxies included.} 
\label{fig:xuv_freq_1&3} 
\end{figure*}

Fig. \ref{fig:xuv_freq_1&3} shows the same sequence of histograms, but also includes in the XUV fraction the number of XUV-ambiguous galaxies.  The peak in the XUV fraction histogram is spread in stellar mass.  The total space density of XUV-disk galaxies is $>$ 1.5-4.2 $\times$ 10$^{-3}$ Mpc$^{-3}$, with the range given by the space density computed for XUV-disks only and for both XUV-disk and XUV-ambiguous galaxies.  We emphasize that this is a lower limit because of the redshift effects discussed above.

\subsection{The Origin of the XUV Emission}
\label{sec:origin}

\subsubsection{Formation Scenarios}
\label{sec:formation}

A number of physical mechanisms including interactions, perturbations, gas accretion, and the outward propagation of spiral density waves have been proposed as triggers of star formation in XUV-disks \citep[see T07;][for a more complete discussion of possible formation scenarios]{Bush08,Bush2010}.  As mentioned in Section \ref{sec:our scheme}, we leave open the possibility that the extended star formation in some XUV-ambiguous galaxies is not distinct from the main body of the galaxies and so does not have a distinct formation scenario.  

Much has been said about the role of interactions in XUV-disks because many of the known XUV-disks show evidence of a recent interaction.  One of the first XUV-disks discovered is considered a direct result of a previously-confirmed interaction \citep[NGC 4625;][]{GilDePaz05}.  \citet{Thilker2010} provides an example of an XUV-disk galaxy following a merger that is undergoing enough star formation to rejuvenate the disk and shift a red lenticular galaxy towards the green valley.  \citetalias{Thilker07} report that 75\% of the galaxies they identified as Type 1 and about 50\% of the galaxies they identified as Type 2 show evidence of an interaction based on the tidal perturbation parameter from \citet{Varela04}.  We leave an analysis of interactions and their relationship to XUV-disks to a later study.

In disk galaxies, it is plausible that the gas dynamics and structuring in the inner disk play a role in generating the star formation in the outer disk.  \citet{Bush08, Bush2010} completed models to determine how XUV emission could be created in disk galaxies.  They showed that in disk galaxies with extended gas disks, the spiral density wave can propagate into the outer disk, producing gravitationally unstable regions that collapse and form stars.  This appears to be a reasonable explanation for the XUV emission in many late-type galaxies even though it doesn't address the origin of the extended gas distribution itself.  It is worth noting that \citet{Bush2010} were successful in reproducing Type 1 XUV-disks but were unable to recreate Type 2 XUV-disks.  They suggest that recent gas accretion might be responsible for producing young, blue Type 2 XUV-disks.  
  
\citetalias{Thilker07} also proposed that gas has an important role in the formation of XUV-disks.  They showed that XUV-disks are twice as gas-rich as would be expected based on their Hubble type.  An enhanced HI content has been linked to a strong color gradient in which the outer disk of a galaxy is much bluer than the inner disk \citep{Wang2010}.  Such galaxies, which are likely undergoing inside-out disk growth, may have experienced a recent gas accretion event which elevated the HI content.  Similar color gradients for the XUV-disks in our sample (see Fig. \ref{fig:colorcut}) point to the possibility that XUV-disks are the result of gas accretion.

Whether the stars in XUV-disks are formed from recently accreted gas or an older gas reservoir, it is important to consider the gas from which these stars form.  Galaxies form and evolve by accreting gas from their surroundings.  The physical processes by which galaxies accrete enough gas to match the measured star formation rate are a subject of active study.  The classical view of gas accretion focused on hot-mode accretion in which infalling gas is shock-heated to the virial temperature and then radiatively cools to form dense clouds that eventually produce stars.  Recently, \citet{Keres09} showed that it is not hot-mode accretion but rather cold-mode accretion that supplies galaxies with most of their fuel for star formation via filaments that are accreted directly from the intergalactic medium (IGM).  \citet{KeresHern09} showed that these filaments can condense into clouds that can then rain onto galaxies and provide gas for star formation. The newly accreted clouds will be distributed in a flattened disk around the galaxy.  These clouds may be analogs of high-velocity clouds (HVCs) surrounding the Milky Way which are thought to provide much of the fuel for star formation in the Galaxy.  \citet{KeresHern09} estimate that, for a Milky Way-sized halo, gas can be supplied in this manner at a rate of 0.6 M$_{\odot}$ yr$^{-1}$.  If gas recycled through the galactic fountain is included, the accretion rate may reach 1 M$_{\odot}$ yr$^{-1}$, providing enough gas to sustain the current star formation rate of 1 M$_{\odot}$ yr$^{-1}$.
  
Attempts to detect accreting gas rely on observational signatures of gas accretion that include HI tails and filaments, accompanying dwarfs, extended and warped HI morphologies, lopsided disks, and extraplanar HI in spiral galaxies \citep{Sancisi08}.  In studies done thus far, measured accretion rates based on such signatures do not account for total star formation rates around 1 M$_{\odot}$ yr$^{-1}$.  Sancisi et al. (2008) find an observed accretion rate of 0.2 M$_{\odot}$ yr$^{-1}$.  \citet{Fraternali2010} cites a number of studies based on measurements of extraplanar neutral gas that report accretion rates accounting for 5-24\% of the star formation rate in each galaxy.  

Recent cosmological simulations done by \citet{Roskar2010} focused on the role of gas accretion in the creation of warped galactic disks and the misalignment of the inner and outer disks that underlie the warp.  They found that cold gas accretion accounts for 75\% of the mass in the misaligned outer disk, and cold gas that is not shock-heated is responsible for much of the star formation in the misaligned outer disk.  They also find that such warps created by the accretion of cold gas onto a misaligned outer disk will disappear in the absence of continuous accretion of cold gas.  Some of the first XUV-disks that were discovered were characterized by warped HI disks (Thilker et al. 2005; T07).  In our sample, at least one galaxy appears to have a slight warp in its XUV-disk (see Section \ref{sec:examples}).  Thus, it is reasonable to assume that there is a strong connection between XUV-disks and cold gas accretion and that XUV-disks may be fueled by cold gas accretion.  \citet{Roskar2010} suggest that observations of stellar populations beyond the central stellar disk, such as those that form in an XUV-disk, will be valuable in probing cold accretion.

\subsubsection{Gas Accretion Rate for Our Sample}
\label{sec:gasaccretion}

Here we use the extended star formation in XUV-disk galaxies as evidence of recent or ongoing gas accretion.  Although XUV-disks cannot be linked unequivocally to gas accretion, we explore the possibility here.  Many of the defining features of XUV-disks, such as rings, are suggestive of recent gas accretion.  We use the calculated XUV-disk space density and the UV properties of XUV-disks to constrain the rate of cold gas accretion onto these galaxies.  The cold gas accretion rate we estimate may include low-redshift cold accretion and cold gas clouds acquired from minor mergers or interactions.  

As discussed above, this work is an exploratory analysis and we caution that all numbers reported here are coarse estimates and the result of several assumptions.  Our main goal here is to set up a framework for future analysis of XUV-disks and observations of gas accretion.  Because our ability to detect faint UV features at moderate redshifts with existing and future technology surpasses our ability to directly measure the HI content of galaxies beyond the very local universe, methods similar to that which we develop here, in which estimates of cold gas accretion are based on the UV properties of galaxies, will play a crucial role in the interpretation of future observations of galaxies.

We use the following equation to determine the gas accretion rate onto XUV-disk galaxies in bins of NUV-r and M$_*$

\begin{equation}
\dot M_{gas,XUV} =\frac{\phi \times f_{xuv} \times <M_{gas,XUV}>}{\Delta T}
\end{equation}

where $\phi$ = $\phi(NUV-r,M_*)$ is the galaxy volume density \citep[Mpc$^{-3}$; derived from][]{Wyder07}, $f_{XUV}$ = $f_{XUV}(NUV-r,M_*)$ is the  volume-averaged XUV fraction calculated above, $M_{gas,XUV}$ = $M_{gas,XUV}(NUV-r,M_*)$ is the average HI mass within the XUV regions, and $\Delta T$ is the accretion timescale.  Below we describe some of the assumptions that went into our calculation.  

To estimate $M_{gas,XUV}$ we adopt an average HI surface density of 3 M$_{\odot}$ pc$^{-2}$ over the deprojected surface area of the XUV regions.  Although this value is somewhat arbitrary and is uncertain by about a factor of 2, we selected it to match that observed for the outer parts of XUV-disks M 83 \citep{Thilker05} and NGC 4625 \citep{GilDePaz05}.  This value is also consistent with the HI detected in outer regions of galaxies by the THINGS survey \citep{Bigiel08}.  For our sample, the average HI mass associated with XUV regions around XUV-disks ranges from 2 $\times$ 10$^{8}$ to 8 $\times$ 10$^{9}$ M$_{\odot}$ with a median value of 2 $\times$ 10$^{9}$ M$_{\odot}$.  

A critical assumption in determining the gas accretion rate is selecting the timescale over which gas is accreted.  In general, the gas consumption timescale in the outer parts of disks is longer than the likely timescale for inflow \citep{Bigiel2010}.  Thus we use a dynamical timescale for the gas accretion timescale ($\Delta T$), which we assume to be 1 Gyr across all bins.  It is not clear that this is the best estimate for each galaxy in our sample, but this is a reasonable approximation based on recent work.  \citet{Sancisi08} show that recently accreted asymmetric features, such as tails, will be incorporated into the parent galaxy over a few dynamical times, or roughly 1 Gyr.  \citet{Haan09} estimate that gravitational torques in spiral galaxies will redistribute cold gas on a similar timescale.  The accretion timescale for the early-type galaxies in our sample may be longer than these estimates, but we use 1 Gyr since it is based on calculations from the literature.

Here we summarize the key assumptions that were made in determining the gas accretion rate that follows:

\begin{enumerate}
\item{The presence of an XUV-disk is indicative of recently accreted HI.}
\item{The XUV region traces the extent of the HI reservoir.}
\item{The HI surface density in the XUV region is consistent with the HI surface density in the outer regions of other well-studied XUV-disks: $\Sigma_{HI}$ = 3 M$_{\odot}$ pc$^{-2}$ .}
\item{The timescale over which the cold gas is accreted is a few dynamical times: $\Delta T$ = 1 Gyr.}
\end{enumerate}

The total amount of cold gas accreting onto XUV-disk galaxies is $>$ 1.7-4.6 $\times$ 10$^6$ M$_{\odot}$ Mpc$^{-3}$, with the lower estimate derived using only XUV-disks and the upper estimate derived using both XUV-disks and XUV-ambiguous galaxies.  The HI volume density of the local universe is 6 $\times$ 10$^7$ M$_{\odot}$ Mpc$^{-3}$ \citep{Zwaan05}.  From this we can estimate that 3-8\% of the HI gas in the universe might be associated with XUV-disks.  

\begin{figure}
\epsscale{1.2}
\plotone{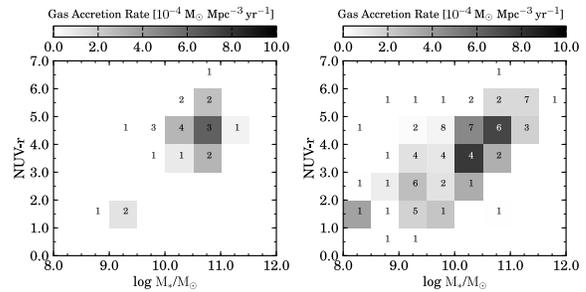}
\caption{Histograms showing the mass of gas accreted onto XUV-disks per bin.  XUV-disks only are included in the plot on the left and XUV-disks and XUV-ambiguous galaxies are included in the plot on the right.  The numbers overlaid on the plots indicate the total number of XUV-disks and XUV-disks plus XUV-ambiguous galaxies in each bin.} 
\label{fig:gas_accr} 
\end{figure}

Our result for the gas accretion rate onto XUV-disk galaxies, in terms of M$_{\odot}$ Mpc$^{-3}$ yr$^{-1}$, is shown in Fig. \ref{fig:gas_accr}.  The infall rate onto XUV-disks is $>$ 1.7-4.6 $\times$ 10$^{-3}$ M$_{\odot}$ Mpc$^{-3}$ yr$^{-1}$ with the range given by the difference between the two panels.  Galaxies throughout the sequence, including those in the red sequence and green valley, are undergoing gas accretion.  The local star formation rate density is 0.02 M$_{\odot}$ Mpc$^{-3}$ yr$^{-1}$ \citep{Salim07}.  If all of the gas in the XUV region eventually forms stars, our liberal estimate that includes XUV-ambiguous galaxies suggests that cold gas accretion onto XUV-disk galaxies could account for about 23\% of the local star formation rate density.  The calculation of the cold gas accretion rate derived from our conservative estimate, which excludes XUV-ambiguous galaxies, only accounts for about 9\% of the star formation in the local universe.  This is consistent with other estimates of the gas accretion rate noted above.  Although simulations \citep[e.g. those of][]{KeresHern09} have shown that gas accretion provides most of the fuel for continued star formation in galaxies, observational estimates of gas accretion rates have consistently underestimated the gas accretion rate by a factor of $\sim$ 5 to 20.

The above calculation provides an estimate of the HI associated with XUV emission in our sample, but of course it cannot speak to the amount of HI accreting onto galaxies that do not exhibit XUV flux.  Although it is reasonable to assume that the galaxies in our sample with XUV-disks have extended HI disks beyond the optical radius, the existence of extended HI in galaxies without XUV-disks is less clear.  Indeed, \citet{Sancisi08} describe the tenuous correlation between gas accretion and star formation rate, as evidenced by the existence of gas-rich ellipticals with no signs of recent star formation.  Thus, our estimate of the amount of gas associated with XUV-disks underestimates the total amount of gas associated with extended HI disks because our methods only allow us to measure the HI associated with disks that support extended star formation.  

\subsubsection{Redshift Correction}
\label{sec:redshift_correction}

\begin{figure}
\epsscale{1.2}
\plotone{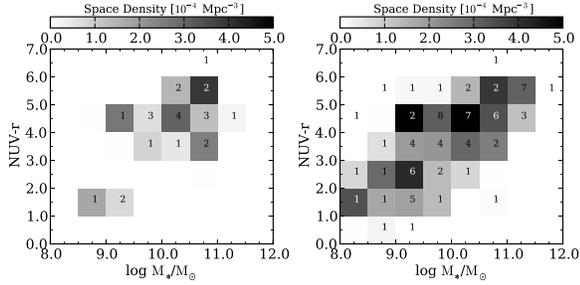}
\caption{Redshift-corrected space density of XUV-disks on the left and XUV-disks plus XUV-ambiguous galaxies on the right.  The numbers overlaid on the plots indicate the total number of XUV-disks and XUV-disks plus XUV-ambiguous galaxies in each bin in the original sample.} 
\label{fig:xuv_freq_red} 
\end{figure}

\begin{figure}
\epsscale{1.2}
\plotone{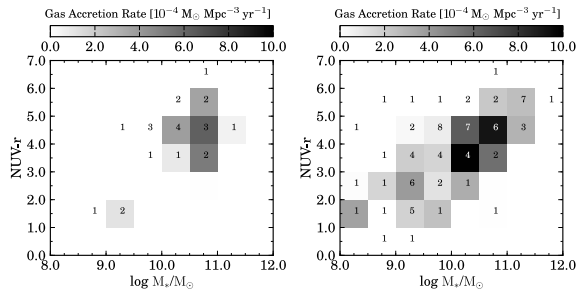}
\caption{Redshift-corrected gas accretion rate onto XUV-disks on the left and onto XUV-disks plus XUV-ambiguous galaxies on the right.  The numbers overlaid on the plots indicate the total number of XUV-disks and XUV-disks plus XUV-ambiguous galaxies in each bin in the original sample.} 
\label{fig:gas_accr_red} 
\end{figure}

As stated in section \ref{sec:redshift}, our ability to detect XUV-disks beyond 100 Mpc is limited by the resolution of \emph{GALEX}.  Here we correct for this limitation by increasing the XUV fractions in each redshift bin to 20\% to match the XUV fraction in the lower redshift bins.  Correcting for incompleteness with respect to redshift, our estimate of the XUV-disk space density is $>$ 2.0-5.9 $\times$ 10$^{-3}$ Mpc$^{-3}$ (see Fig. \ref{fig:xuv_freq_red}), a modest increase over the space density determined without the correction.  The gas accretion rate onto such galaxies increases to $>$ 2.1-6.2 $\times$ 10$^{-3}$ M$_{\odot}$ Mpc$^{-3}$ yr$^{-1}$ (see Fig. \ref{fig:gas_accr_red}).

\subsection{XUV-disks and Their Relation to Disk-Building}
\label{sec:disk-building}

Given that the gas accretion rate is significant ($>$ 1.3-2.8 $\times$ 10$^{-3}$ M$_{\odot}$ Mpc$^{-3}$ yr$^{-3}$) even for XUV-disks around transition galaxies in the green valley, it is reasonable to consider that the infall of gas onto the XUV-disks lying in the green valley may lead to enough star formation to cause the galaxy to transition away from the red sequence.  Our results show that the XUV-disk fraction in the green valley is 7-18\%; thus, it is possible that a similar fraction of green valley galaxies might be moving away from the red sequence due to a new round of star formation in their outer regions.

Previous studies of the green valley \citep[e.g.][]{Martin07} focused on the quenching processes occurring in green valley galaxies that might drive galaxies from the blue sequence to the red sequence.  More recently, work has been done to target transition galaxies in the green valley in order to identify galaxies that are transitioning either to or from the red sequence \citep[e.g.][]{Catinella2010}.  For example, NGC 404 is a lenticular galaxy that may be transitioning away from the red sequence.  \citet{Thilker2010} showed that if its HI ring is the result of a recent accretion event \citep[as proposed by][]{delRio04}, the star formation in the ring caused the galaxy to move from the red sequence into the green valley.  

More recently, \citet{Williams2010} used HST observations of the galaxy to show that its star formation rate has since decreased after a brief increase immediately following the accretion event.  Thus, although the accretion of new material initially led to an increased level of star formation that caused the galaxy to transition to the green valley, NGC 404 will likely fall back to the red sequence instead of moving blueward.  NGC 404 is not a lone case: the majority of the massive early-type galaxies in the sample studied by \citet{Salim2010} appear to be similar to NGC 404.  They, too, lie in the green valley and have rejuvenated disks that are probably the result of gas accretion from the IGM or minor mergers.

The temporary detour to the green valley that was seen for NGC 404 provides support for our suggestion that XUV-disks represent a phase in the evolution of a galaxy.  There are numerous examples in the literature of galaxies whose morphologies are indicative of XUV-disks and that appear to be undergoing a star formation event that is associated with gas accretion.  Indeed, \citet{Cortese09} discovered a population of transition galaxies with a ``normal"  amount of HI (most of the transition galaxies they studied were HI-deficient) that are moving away from the red sequence as the HI is consumed and converted to stars.  Many of those HI-normal transition galaxies with obvious recent star formation display prominent UV rings.  The presence of the UV rings suggests that the population \citet{Cortese09} describe includes XUV-disk galaxies that are in the process of disk-building.

\citet{Moffett09} interpreted XUV-disks in E/S0s as evidence for disk-building following a merger and \citet{Kannappan09} suggested that blue E/S0s, which may make up a fraction of our XUV-disks, are probably building disks as well.  That disk-building can be a result of gas accretion is consistent with \citet{KeresHern09} - they suggest that the cold gas accretion will form a disk around an evolved galaxy, providing fuel for extended star formation and a path away from the red sequence.

At this point, it is not easy to distinguish between transition galaxies that are experiencing a new wave of gas accretion and subsequently moving away from the red sequence and transition galaxies whose gas accretion rate is slowly decreasing, causing them to move towards the red sequence.  Ring-like features around early-type galaxies are often interpreted as signs of recent gas accretion, but it is possible that the phase of gas accretion represented by the ring is in fact ending.  Analysis of a galaxy's potential for new and continued star formation (and subsequent movement within the color-magnitude diagram) is locked up in the amount of HI gas in the galaxy.  Such a detailed analysis is beyond the scope of this paper but crucial in determining exactly how much star formation can be expected in the XUV-disk and what effects it may have on galaxy morphology.  Future studies should focus on trying to disentangle observations of gas accretion and quenching of star formation so that we are able to conclusively identify a population of galaxies that are re-building disks and explore its connection to the population of XUV-disks.

\section{CONCLUSIONS}
\label{sec:conclusions}

We report the XUV-disk space density in the local universe based on an unbiased survey of galaxies in the intersection of available GALEX deep imaging and SDSS footprints.  Galaxies of all colors and masses exhibit XUV-disks, but a higher fraction of red massive early-type galaxies show evidence of star formation in their outer extents.  We investigated the possibility that the extended star formation in many XUV-disk galaxies is due to cold mode accretion by estimating the gas accretion rate onto XUV-disks.  There is a significant level of gas accretion onto all galaxy types, including transition galaxies in the green valley.  Gas accretion onto galaxies in the green valley may represent evidence that these galaxies are rebuilding their disks.

Some of our key results follow:
\begin{enumerate}
\item{Based on our measurements and simulations, we find that deep GALEX imaging allows us to detect XUV-disks beyond 100 Mpc.}
\item{We have used our unbiased survey to establish the average frequency of XUV-disks out to z=0.05 as 4-14\%, with 4\% as a hard lower limit.  The incidence rises to close to 20\% for the nearby portion of our sample (somewhat consistent with, though lower than, previous findings; i.e. those of \citetalias{Thilker07})}.
\item{The calculated XUV-disk fraction along with measurements of the local space density of galaxies allows us to estimate that the space density of XUV-disks is $>$ 1.5-4.2 $\times$ 10$^{-3}$ Mpc$^{-3}$.  We used an estimate of the gas content associated with XUV-disk galaxies to establish a gas accretion rate onto XUV-disk galaxies of $>$ 1.7-4.6 $\times$ 10$^{-3}$ M$_{\odot}$ $^{-3}$ yr$^{-1}$.}
\item{Under the assumption that XUV-disk galaxies in the green valley might be rebuilding their disks, we find that 7-18\% of galaxies in the green valley could be transitioning away from the red sequence.}
\end{enumerate}

The work presented here represents an attempt to expand the known sample of XUV-disk galaxies and to estimate global properties of such galaxies.  We are limited by the resolution of GALEX and thus restricted in our ability to make firm conclusions about the prevalence of such galaxies in different mass ranges and at various redshifts.  Future searches for XUV-disks will require data with greater sensitivity and finer resolution in order to detect small pockets of star formation in the outer reaches of galaxies.

\acknowledgments

We thank the anonymous referee for valuable comments that substantially improved the quality of this paper.

\emph{GALEX (Galaxy Evolution Explorer)} is a NASA Small Explorer, launched
in April 2003. We gratefully acknowledge NASA's support for
construction, operation, and science analysis for the \emph{GALEX} mission,
developed in cooperation with the Centre National d'Etudes Spatiales
(CNES) of France and the Korean Ministry of Science and Technology.

This work has made extensive use of the \url[]{MPA/JHU} and the \url[]{NYU}
SDSS value-added catalogs.

Funding for the SDSS and SDSS-II has been provided by the Alfred
P. Sloan Foundation, the Participating Institutions, the National
Science Foundation, the U.S. Department of Energy, the National
Aeronautics and Space Administration, the Japanese Monbukagakusho, the
Max Planck Society, and the Higher Education Funding Council for
England. The SDSS Web Site is http://www.sdss.org/.

The SDSS is managed by the Astrophysical Research Consortium for the
Participating Institutions. The Participating Institutions are the
American Museum of Natural History, Astrophysical Institute Potsdam,
University of Basel, University of Cambridge, Case Western Reserve
University, University of Chicago, Drexel University, Fermilab, the
Institute for Advanced Study, the Japan Participation Group, Johns
Hopkins University, the Joint Institute for Nuclear Astrophysics, the
Kavli Institute for Particle Astrophysics and Cosmology, the Korean
Scientist Group, the Chinese Academy of Sciences (LAMOST), Los Alamos
National Laboratory, the Max-Planck-Institute for Astronomy (MPIA),
the Max-Planck-Institute for Astrophysics (MPA), New Mexico State
University, Ohio State University, University of Pittsburgh,
University of Portsmouth, Princeton University, the United States
Naval Observatory, and the University of Washington.

{\it Facilities:} \facility{GALEX}, \facility{Sloan}

\appendix

\section{MODIFICATIONS TO SPACE DENSITY AND GAS ACCRETION RATE ESTIMATES BASED ON COLOR OF XUV REGION}

\begin{figure}
\epsscale{0.8}
\plotone{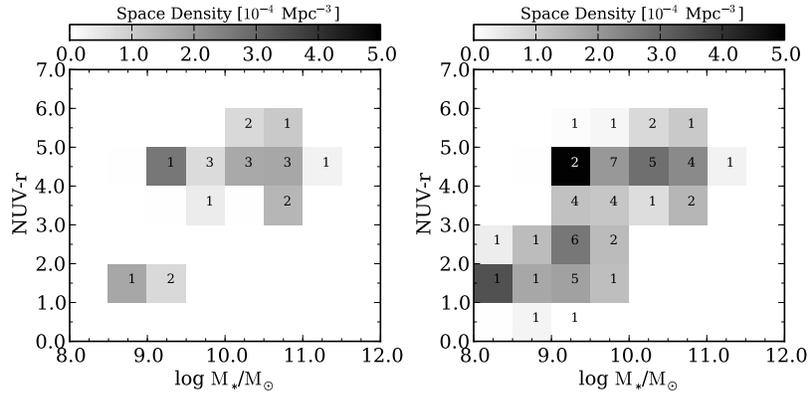}
\caption{Space density of XUV-disks on the left and XUV-disks plus XUV-ambiguous galaxies on the right.  Rejects with FUV-\emph{r}$_{\mathrm{XUV}}$ $>$ 5 have been removed from the sample.  The numbers overlaid on the plots indicate the total number of XUV-disks and XUV-disks plus XUV-ambiguous galaxies in each bin after rejects were removed from the sample.} 
\label{fig:xuv_freq_reject} 
\end{figure}

\begin{figure}
\epsscale{0.8}
\plotone{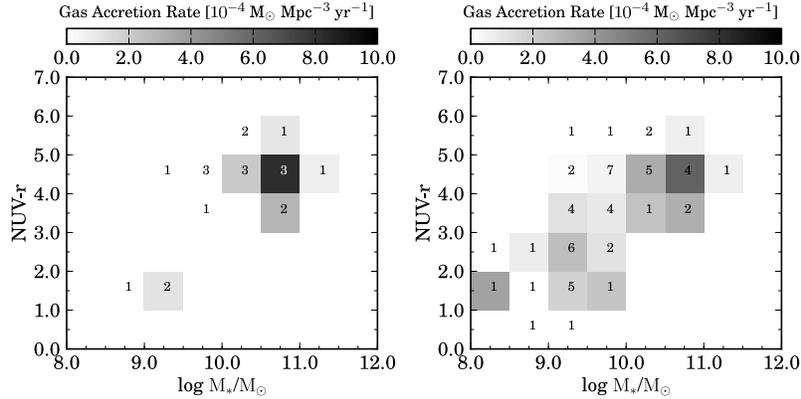}
\caption{Gas accretion rate onto XUV-disks on the left and onto XUV-disks plus XUV-ambiguous galaxies on the right.  Rejects with FUV-\emph{r}$_{\mathrm{XUV}}$ $>$ 5 have been removed from the sample.  The numbers overlaid on the plots indicate the total number of XUV-disks and XUV-disks plus XUV-ambiguous galaxies in each bin after rejects were removed from the sample.} 
\label{fig:gas_accr_reject} 
\end{figure}

In section \ref{sec:color cut} we identified some XUV-disks and XUV-ambiguous galaxies as potential ``rejects" based on the color of their XUV regions suggesting that the extended star formation may not have proceeded recently or is not distinct from the star formation in the inner disk.  Here we exclude ``rejects," 4 XUV-disks and 21 XUV-ambiguous galaxies with FUV-r$_{\mathrm{XUV}}$ $>$ 5.0, from our estimates of XUV-disk frequency and gas accretion rates.  This color cut yields an XUV-disk space density of $>$ 1.2-3.0 $\times$ 10$^{-3}$ Mpc$^{-3}$ (see Fig. \ref{fig:xuv_freq_reject}) and a gas accretion rate onto XUV-disks of $>$ 1.6-3.1 $\times$ 10$^{-3}$ M$_{\odot}$ Mpc$^{-3}$ yr$^{-1}$ (see Fig. \ref{fig:gas_accr_reject}).

 
 \end{document}